\RequirePackage[table]{xcolor}
\documentclass[table]{article}
\usepackage{lmodern}
\usepackage{amssymb,amsmath}
\usepackage{ifxetex,ifluatex}
\ifnum 0\ifxetex 1\fi\ifluatex 1\fi=0 
  \usepackage[T1]{fontenc}
  \usepackage[utf8]{inputenc}
\else 
  \ifxetex
    \usepackage{mathspec}
  \else
    \usepackage{fontspec}
  \fi
  \defaultfontfeatures{Ligatures=TeX,Scale=MatchLowercase}
\fi
\IfFileExists{upquote.sty}{\usepackage{upquote}}{}
\IfFileExists{microtype.sty}{%
\usepackage{microtype}
\UseMicrotypeSet[protrusion]{basicmath} 
}{}
\usepackage[margin=1in]{geometry}
\usepackage{hyperref}
\hypersetup{unicode=true,
            pdftitle={An Introductory Tutorial on Cohort State-Transition Models in R Using a Cost-Effectiveness Analysis Example},
            pdfauthor={Fernando Alarid-Escudero, PhD; Eline Krijkamp, MSc; Eva A. Enns, PhD;  Alan Yang, MSc; M. G. Myriam Hunink, PhD\^{}\textbackslash{}dagger; Petros Pechlivanoglou, PhD; Hawre Jalal, MD, PhD},
            pdfkeywords={Markov models, state-transition models, decision models, Tutorial, R},
            pdfborder={0 0 0},
            breaklinks=true}
\usepackage{color}
\usepackage{fancyvrb}

\DefineVerbatimEnvironment{Highlighting}{Verbatim}{commandchars=\\\{\}}
\usepackage{framed}
\definecolor{shadecolor}{RGB}{248,248,248}
\newenvironment{Shaded}{\begin{snugshade}}{\end{snugshade}}

\newcommand{\DecValTok}[1]{\textcolor[rgb]{0.00,0.00,0.81}{#1}}

\newcommand{\FloatTok}[1]{\textcolor[rgb]{0.00,0.00,0.81}{#1}}
\newcommand{\ConstantTok}[1]{\textcolor[rgb]{0.00,0.00,0.00}{#1}}

\newcommand{\SpecialCharTok}[1]{\textcolor[rgb]{0.00,0.00,0.00}{#1}}
\newcommand{\StringTok}[1]{\textcolor[rgb]{0.31,0.60,0.02}{#1}}

\newcommand{\CommentTok}[1]{\textcolor[rgb]{0.56,0.35,0.01}{\textit{#1}}}
\newcommand{\DocumentationTok}[1]{\textcolor[rgb]{0.56,0.35,0.01}{\textbf{\textit{#1}}}}

\newcommand{\OtherTok}[1]{\textcolor[rgb]{0.56,0.35,0.01}{#1}}
\newcommand{\FunctionTok}[1]{\textcolor[rgb]{0.00,0.00,0.00}{#1}}

\newcommand{\ControlFlowTok}[1]{\textcolor[rgb]{0.13,0.29,0.53}{\textbf{#1}}}

\newcommand{\AttributeTok}[1]{\textcolor[rgb]{0.77,0.63,0.00}{#1}}

\newcommand{\NormalTok}[1]{#1}
\usepackage{longtable,booktabs}
\usepackage{graphicx,grffile}
\makeatletter
\def\maxwidth{\ifdim\Gin@nat@width>\linewidth\linewidth\else\Gin@nat@width\fi}
\def\maxheight{\ifdim\Gin@nat@height>\textheight\textheight\else\Gin@nat@height\fi}
\makeatother
\setkeys{Gin}{width=\maxwidth,height=\maxheight,keepaspectratio}
\IfFileExists{parskip.sty}{%
\usepackage{parskip}
}{
\setlength{\parindent}{0pt}
\setlength{\parskip}{6pt plus 2pt minus 1pt}
}
\ifx\paragraph\undefined\else
\let\oldparagraph\paragraph
\renewcommand{\paragraph}[1]{\oldparagraph{#1}\mbox{}}
\fi
\ifx\subparagraph\undefined\else
\let\oldsubparagraph\subparagraph
\renewcommand{\subparagraph}[1]{\oldsubparagraph{#1}\mbox{}}
\fi

\let\rmarkdownfootnote\footnote%
\def\footnote{\protect\rmarkdownfootnote}

\usepackage{titling}

\newlength{\cslhangindent}
\newlength{\csllabelwidth}
\newenvironment{CSLReferences}[2] 
 {
  \setlength{\parindent}{0pt}
  \ifodd #1 \everypar{\setlength{\hangindent}{\cslhangindent}}\ignorespaces\fi
  \ifnum #2 > 0
  \setlength{\parskip}{#2\baselineskip}
  \fi
 }%
 {}
\usepackage{calc}

\newcommand{\CSLLeftMargin}[1]{\parbox[t]{\csllabelwidth}{#1}}
\newcommand{\CSLRightInline}[1]{\parbox[t]{\linewidth - \csllabelwidth}{#1}\break}

  \title{An Introductory Tutorial on Cohort State-Transition Models in R Using a Cost-Effectiveness Analysis Example}
    \pretitle{\vspace{\droptitle}\centering\huge}
  \posttitle{\par}
    \author{Fernando Alarid-Escudero, PhD\footnote{Division of Public Administration, Center for
  Research and Teaching in Economics (CIDE), Aguascalientes,
  Aguascalientes, Mexico} \\ Eline Krijkamp, MSc\footnote{Department of Epidemiology and Department of Radiology,  Erasmus
  University Medical Center, Rotterdam, The Netherlands} \\ Eva A. Enns, PhD\footnote{Division of Health Policy and Management,
  University of Minnesota School of Public Health, Minneapolis, MN, USA} \\ Alan Yang, MSc\footnote{The Hospital for Sick Children,
  Toronto, Ontario, Canada} \\  M.G. Myriam Hunink, PhD\(^\dagger\)\footnote{Center for Health Decision
  Sciences, Harvard T.H. Chan School of Public Health, Boston, USA} \\ Petros Pechlivanoglou, PhD\footnote{The Hospital for Sick Children,
  Toronto and University of Toronto, Toronto, Ontario, Canada} \\ Hawre Jalal, MD, PhD\footnote{School of Epidemiology and Public Health, Faculty of Medicine, University of Ottawa, Ottawa, ON, Canada}}
    \preauthor{\centering\large\emph}
  \postauthor{\par}
      \predate{\centering\large\emph}
  \postdate{\par}
    \date{\today}

\usepackage{booktabs}
\usepackage{longtable}
\usepackage{array}
\usepackage{multirow}
\usepackage{wrapfig}
\usepackage{float}
\usepackage{colortbl}
\usepackage{pdflscape}
\usepackage{tabu}
\usepackage{threeparttable}
\usepackage{threeparttablex}
\usepackage[normalem]{ulem}
\usepackage{makecell}

\usepackage{amsmath}
\usepackage{float}
\usepackage{setspace}

\begin{document}
\maketitle
\begin{abstract}
Decision models can combine information from different sources to simulate the long-term consequences of alternative strategies in the presence of uncertainty. A cohort state-transition model (cSTM) is a decision model commonly used in medical decision-making to simulate the transitions of a hypothetical cohort  among various health states over time. This tutorial focuses on time-independent cSTM, where transition probabilities among health states remain constant over time. We implement time-independent cSTM in R, an open-source mathematical and statistical programming language. We illustrate time-independent cSTMs using a previously published decision model, calculate costs and effectiveness outcomes, conduct a cost-effectiveness analysis of multiple strategies, including a probabilistic sensitivity analysis. We provide open-source code in R to facilitate wider adoption. In a second more advanced tutorial, we illustrate time-dependent cSTMs.

\end{abstract}

\section{Introduction}\label{introduction}

Policymakers are often tasked with allocating limited healthcare resources under constrained budgets and uncertainty about future outcomes. Health economic evaluations might inform their final decisions. These economic evaluations often rely on decision models to synthesize evidence from different sources and project long-term outcomes of various alternative strategies. A commonly used decision model is the discrete-time cohort state-transition model (cSTM), often referred to as a Markov model.\textsuperscript{\protect\hyperlink{ref-Kuntz2017}{1}}

A cSTM is a dynamic mathematical model in which a hypothetical cohort of individuals transition between different health states over time. A cSTM is most appropriate when the decision problem has a dynamic component (e.g., the disease process can vary over time) and can be described using a reasonable number of health states. cSTMs are often used because of their transparency, efficiency, ease of development, and debugging. cSTMs are usually computationally less demanding than iSTMs, providing the ability to conduct PSA and value-of-information (VOI) analyses that otherwise might not be computationally feasible with iSTMs.\textsuperscript{\protect\hyperlink{ref-Siebert2012c}{2}} cSTMs have been used to evaluate screening and surveillance programs,\textsuperscript{\protect\hyperlink{ref-Suijkerbuijk2018}{3},\protect\hyperlink{ref-Sathianathen2018a}{4}} diagnostic procedures,\textsuperscript{\protect\hyperlink{ref-Lu2018b}{5}} disease management programs,\textsuperscript{\protect\hyperlink{ref-Djatche2018}{6}} interventions,\textsuperscript{\protect\hyperlink{ref-Smith-Spangler2010}{7}} and policies.\textsuperscript{\protect\hyperlink{ref-Pershing2014}{8}}

In a recent review, we illustrated the increased use of R's statistical programming framework in health decision sciences. We provided a summary of available resources to apply to medical decision making.\textsuperscript{\protect\hyperlink{ref-Jalal2017b}{9}} Many packages have been explicitly developed to estimate and construct cSTMs in R. For example, the \texttt{markovchain}\textsuperscript{\protect\hyperlink{ref-Spedicato2017}{10}} and \texttt{heemod}\textsuperscript{\protect\hyperlink{ref-Filipovic-Pierucci2017}{11}} packages are designed to build cSTMs using a pre-defined structure. The \texttt{markovchain} package simulates time-independent, and time-dependent Markov chains but is not designed to conduct economic evaluations. \texttt{heemod} is a well-structured R package for economic evaluations. However, these packages are necessarily stylized and require users to specify the structure and inputs of their cSTM in a particular way potentially without fully understanding how cSTMs work. Using these packages can be challenging if the desired cSTM does not fit within this structure.

This tutorial demonstrates how to conduct a full cost-effectiveness analysis (CEA) comparing multiple interventions and implementing probabilistic sensitivity analysis (PSA) without needing a specialized cSTM package. We first describe each of the components of a time-independent cSTM. Then, we illustrate the implementation of these components with an example. Our general conceptualization should apply to other programming languages (e.g., MATLAB, Python, C++, and Julia). The reader can find the most up-to-date R code of the time-independent cSTM and the R code to create the tutorial graphs in the accompanying GitHub repository (\url{https://github.com/DARTH-git/cohort-modeling-tutorial-intro}) to replicate and modify the example to fit their needs. We assume that the reader is familiar with the basics of decision modeling and has a basic understanding of programming. Thus, a prior introduction to R, ``for'' loops, and linear algebra for decision modelers is recommended. The linear algebra concepts used throughout the code are explained in more detail in the Supplementary Material.

This introductory tutorial aims to (1) conceptualize time-independent cSTMs for implementation in a programming language and (2) provide a template for implementing these cSTMs in \emph{base} R. We focus on using R \emph{base} packages, ensuring modelers understand the concept and structure of cSTMs and avoid the limitation of constructing cSTMs in a package-specific structure. We used previously developed R packages for visualizing CEA results and checking cSTMs are correctly specified.

\section{Cohort state-transition models (cSTMs)}\label{cohort-state-transition-models}

A cSTM consists of a set of \(n_S\) mutually exclusive and collectively exhaustive health states. The cohort is assumed to be homogeneous within each health state. Individuals in the cohort residing in a particular health state are assumed to have the same characteristics and are indistinguishable from one another. The cohort could transition between health states with defined probabilities, which are called ``transition probabilities''. A transition probability represents the chance that individuals in the cohort residing in a state in a given cycle transition to another state or remain in the same state. In a cSTM, a one-cycle transition probability reflects a conditional probability of transitioning during the cycle, given that the person is alive at the beginning of the cycle.\textsuperscript{\protect\hyperlink{ref-Miller1994}{12}}

In a cSTM, the transition probabilities only depend on the current health state in a given cycle, meaning that the transition probabilities do not depend on the history of past transitions or time spent in a given state. This property is often referred to as the ``Markovian assumption.''\textsuperscript{\protect\hyperlink{ref-Kuntz2001}{13}--\protect\hyperlink{ref-Beck1983}{15}} This tutorial focuses on time-independent cSTM, meaning that model parameters, such as transition probabilities or reward (e.g., costs or utilities associated with being in a particular health state), do not vary with time. We discuss time-dependence in cSTMs in an accompanying advanced tutorial.\textsuperscript{\protect\hyperlink{ref-Alarid-Escudero2021b}{16}}

\hypertarget{rates-versus-probabilities}{%
\subsection{Rates versus probabilities}\label{rates-versus-probabilities}}

In discrete-time cSTMs, cohort dynamics are described by the probability of transitioning between states. However, these transitions might be reported in terms of rates in the literature, or probabilities may not always be available in the desired cycle length. For example, transition probabilities might be available from published literature in one time period (e.g., annual) and might differ from the model's cycle length scale (e.g., monthly). Below, we illustrate a simple approach to converting from rates to probabilities and using rates to convert probabilities from one time scale to another.

While probabilities and rates are often numerically similar in practice, there is a subtle but important conceptual difference between them. A rate represents the \textit{instantaneous} force of an event occurrence per unit time, while a probability represents the cumulative risk of an event over a defined period.

To illustrate this difference further, let us assume that after 10,000 person-years of observation of healthy individuals (e.g., 10,000 individuals observed for an average of 1 year, or 5,000 individuals observed for an average of 2 years, etc.), we observe 500 events of interest (e.g., becoming sick from some disease). The annual event rate of becoming sick, \(\mu_{yearly}\), is then equal to \(\mu_{yearly}=500 / 10,000=0.05\).

If we then wanted to know what proportion of an initially healthy cohort becomes sick at the end of the year, we can convert the annual rate of becoming sick into an annual probability of becoming sick using the following equation:
\begin{equation}
    p_{yearly} = 1-\exp{\left(-\mu_{yearly} \right)},
    \label{eq:rate-to-prob-ann}
\end{equation}
resulting in \(p_{yearly} = 1-\exp{(-0.05)}=\) 0.0488. Equation \eqref{eq:rate-to-prob-ann} assumes that the rate of becoming sick is constant over the year, implying that the time until a healthy person becomes sick is exponentially distributed. The parameter \(p_{yearly}\) is the transition probability from healthy to sick in a cSTM when using an annual cycle length.

If we were concerned that an annual cycle length was too long to capture disease dynamics accurately, we could use a monthly cycle length. To calculate monthly rates, we divide the annual rate by 12:
\begin{equation}
    \mu_{monthly} = \mu_{yearly} / 12.
    \label{eq:rate-ann-to-month}
\end{equation}
To convert to monthly transition probabilities, we apply equation \eqref{eq:rate-to-prob-ann}:
\begin{equation}
    p_{monthly} = 1-\exp{\left(-\mu_{monthly}\right)}.
\end{equation}

We divide by 12 because of the number of months (desired cycle length) in a year (cycle length of the given data). If the original or desired cycle length differed, we would divide by a different factor (e.g., annual to weekly: 52; monthly to annual: 1/12; annual to daily: 365.25, etc.). 

These equations are also helpful for computing probabilities when studies (e.g., survival analyses) provide rates rather than transition probabilities assuming exponentially distributed transition times.

\hypertarget{time-independent-cstm-dynamics}{%
\section{Time-independent cSTM dynamics}\label{time-independent-cstm-dynamics}}

A cSTM consists of three core components: (1) a state vector, \(\mathbf{m}_t\), that stores the distribution of the cohort across all health states in cycle \(t\) where \(t = 0,\ldots, n_T\); (2) the cohort trace matrix, \(M\), that stacks \(\mathbf{m}_t\) for all \(t\) and represents the distribution of the cohort in the various states over time; and (3) a transition probability matrix, \(P\).\textsuperscript{\protect\hyperlink{ref-Iskandar2018a}{17}} If the cSTM is comprised of \(n_S\) discrete health states, \(\mathbf{m}_t\) is a \(1 \times n_S\) vector and \(P\) is a \(n_S \times n_S\) matrix. The \(i\)-th element of \(\mathbf{m}_t\), where \(i = 1,\ldots, n_S\), represents the proportion of the cohort in the \(i\)-th health state in cycle \(t\), referred to as \(m_{[t,i]}\). Thus, \(\mathbf{m}_t\) is written as:
\[
\mathbf{m}_t =
  \begin{bmatrix}
m_{[t,1]} & m_{[t,2]} & \cdots & m_{[t,n_S]}
\end{bmatrix}.
\]
The elements of \(P\) are the transition probabilities of moving from state \(i\) to state \(j\), \(p_{[i,j]}\), where \(\{i,j\} = 1,\ldots, n_S\) and all should have values between 0 and 1.
\[
  P = 
  \begin{bmatrix}
    p_{[1,1]} & p_{[1,2]} & \cdots & p_{[1,n_S]} \\
    p_{[2,1]} & p_{[2,2]} & \cdots & p_{[2,n_S]} \\
    \vdots    & \vdots  & \ddots & \vdots   \\
    p_{[n_S,1]} & p_{[n_S,2]} & \cdots & p_{[n_S,n_S]} \\
  \end{bmatrix}.
\]
For \(P\) to be a correctly specified transition probability matrix, each row of the transition probability matrix must sum to one, \(\sum_{j=1}^{n_S}{p_{[i,j]}} = 1\) for all \(i = 1,\ldots,n_S\).

The state vector at cycle \(t+1\) (\(\mathbf{m}_{t+1}\)) is then calculated as the matrix product of the state vector at cycle \(t\), \(\mathbf{m}_{t}\), and the transition probability matrix, \(P\), such that

\[
  \mathbf{m}_{t+1} = \mathbf{m}_{t} P \text{ for } t = 0,\ldots, (n_T - 1),
\]
where \(\mathbf{m}_1\) is computed from \(\mathbf{m}_{0}\), the initial state vector with the distribution of the cohort across all health states at the start of the simulation (cycle 0). Then, we iteratively apply this equation through \(t = \left(n_T-1 \right)\).

The cohort trace matrix, \(M\), is a matrix of dimensions \((n_T+1) \times n_S\) where each row is a state vector \((-\mathbf{m}_{t}-)\), such that

\[
  M = 
  \begin{bmatrix}
    - \mathbf{m}_0 -  \\
    - \mathbf{m}_1 -  \\
     \vdots \\
    - \mathbf{m}_{n_T} -  
  \end{bmatrix}. 
\]

Note that the initial cycle (i.e., cycle 0) corresponds to \(t=0\), which is on the first row of \(M\). Thus, \(M\) stores the output of the cSTM, which could be used to compute various epidemiological, and economic outcomes, such as life expectancy, prevalence, cumulative resource use, and costs, etc. Table \ref{tab:cSTM-components-table} describes the elements related to the core components of cSTM and their suggested R code names. For a more detailed description of the variable types, data structure, R name for all cSTM elements, please see the Supplementary Material.

\begin{longtable}[]{@{}llcl@{}}
\caption{\label{tab:cSTM-components-table} Components of a cSTM with their R name.}\tabularnewline
\toprule
Element & Description & R name & \\
\midrule
\endfirsthead
\toprule
Element & Description & R name & \\
\midrule
\endhead
\(n_S\) & Number of states & \texttt{n\_states} & \\
\(\mathbf{m}_0\) & Initial state vector & \texttt{v\_m\_init} & \\
\(\mathbf{m}_t\) & State vector in cycle \(t\) & \texttt{v\_mt} & \\
\(M\) & Cohort trace matrix & \texttt{m\_M} & \\
\(P\) & Time-independent transition probability matrix & \texttt{m\_P} & \\
\bottomrule
\end{longtable}

\hypertarget{case-study-sick-sicker-model}{%
\section{Case study: Sick-Sicker model}\label{case-study-sick-sicker-model}}

\begin{figure}[H]

{\centering \includegraphics[width=10.64in]{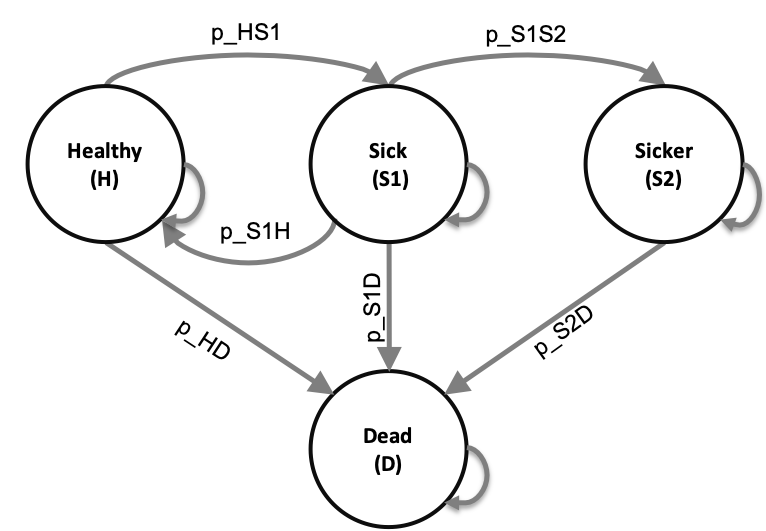} 

}

\caption{State-transition diagram of the time-independent Sick-Sicker cohort state-transition model, showing all possible states (labeled with state names) and transitions (labeled with transition probability variable names).}\label{fig:STD-Sick-Sicker}
\end{figure}

Here, we use the previously published 4-state ``Sick-Sicker'' model for conducting a CEA of multiple strategies to illustrate the various aspects of cSTM implementation in R.\textsuperscript{\protect\hyperlink{ref-Enns2015e}{18}\protect\hyperlink{ref-Krijkamp2018}{19}} Figure \ref{fig:STD-Sick-Sicker} represents the state-transition diagram of the Sick-Sicker model.

The model simulates a cohort at risk of a hypothetical disease with two stages, ``Sick'' and ``Sicker'', to compute the expected costs and quality-adjusted life years (QALYs) of the cohort over time.
All the parameters of the Sick-Sicker model and the corresponding R variable names are presented in Table \ref{tab:param-table}. The naming of these parameters and variables follows the notation described in the DARTH coding framework.\textsuperscript{\protect\hyperlink{ref-Alarid-Escudero2019e}{20}} Briefly, we define variables by \texttt{\textless{}x\textgreater{}\_\textless{}y\textgreater{}\_\textless{}var\_name\textgreater{}}, where \texttt{x} is the prefix that indicates the data type (e.g., scalar (no prefix), \texttt{v} for vector, \texttt{m} for matrix, \texttt{a} for array, \texttt{df} for data frame, etc.), \texttt{y} is the prefix indicating variable type (e.g., \texttt{p} for probability, \texttt{r} for rate, \texttt{hr} for hazard ratio, \texttt{c} for cost \texttt{c}, \texttt{u} for utility, etc.), and \texttt{var\_name} is some description of the variable presented separated by underscores. For example, \texttt{v\_p\_HD} denotes the vector of transition probabilities from health state ``H'' to health state ``D''. In later sections we will define and name all the other parameters.

In this model, we simulate a hypothetical cohort of 25-year-olds in the ``Healthy'' state (denoted ``H'') until they reach a maximum age of 100 years. We will simulate the cohort dynamics in annual cycle lengths, requiring a total of 75 one-year cycles. The total number of cycles is denoted as \(n_T\) and defined in R as \texttt{n\_cycles}. The model setup is as follows. Healthy individuals are at risk of developing the disease when they transition to the ``Sick'' state (denoted by ``S1'') with an annual rate of \texttt{r\_HS1}. Sick individuals are at risk of further progressing to a more severe disease stage, the ``Sicker'' health state (denoted by ``S2'') with an annual rate of \texttt{r\_S1S2}. Individuals in S1 can recover and return to H, as depicted in Figure \ref{fig:STD-Sick-Sicker} by the arc labeled \texttt{p\_S1H}. However, once individuals reach S2, they cannot recover; the rate of transitioning to S1 or H from S2 is zero. Individuals in H face a constant background mortality rate (labeled \texttt{r\_S1H} in Figure \ref{fig:STD-Sick-Sicker}) due to other causes of death. Individuals in S1 and S2 face an increased hazard of death, compared to healthy individuals, in the form of a hazard ratio (HR) of 3 and 10, respectively, relative to the background mortality rate. We transform all transition rates to probabilities following the Section ``Rates versus probabilities'' approach. All transitions between non-death states are assumed to be conditional on surviving each cycle. Individuals in S1 and S2 also experience increased health care costs and reduced quality of life (QoL) compared to individuals in H. When individuals die, they transition to the absorbing ``Dead'' state (denoted by ``D''), meaning that once the proportion of the cohort arrives in that state, they remain. We discount both costs and QALYs at an annual rate of 3\%.

We are interested in evaluating the cost-effectiveness of four strategies: the standard of care (strategy SoC), strategy A, strategy B, and a combination of strategies A and B (strategy AB). Strategy A involves administering treatment A that increases the QoL of individuals in S1 from 0.75 (utility without treatment, \texttt{u\_S1}) to 0.95 (utility with treatment A, \texttt{u\_trtA}). Treatment A costs \$12,000 per year (\texttt{c\_trtA}).\textsuperscript{\protect\hyperlink{ref-Krijkamp2018}{19}} This strategy does not impact the QoL of individuals in S2, nor does it change the risk of becoming sick or progressing through the sick states. Strategy B uses treatment B to reduce only the rate of Sick individuals progressing to the Sicker state by 40\% (i.e., a hazard ratio (HR) of 0.6, \texttt{hr\_S1S2\_trtB}), costs \$13,000 per year (\texttt{c\_trtB}), and does not affect QoL. Strategy AB involves administering both treatments A and B.

We assume that it is not possible to distinguish between Sick and Sicker patients; therefore, individuals in both disease states receive the treatment under the treatment strategies. After comparing the four strategies in terms of expected QALYs and costs, we calculate the incremental cost per QALY gained between non-dominated strategies.

\begin{longtable}[]{@{}lccc@{}}
\caption{\label{tab:param-table} Description of parameters, their R variable name, base-case values and distribution.}\tabularnewline
\toprule
\begin{minipage}[b]{(\columnwidth - 3\tabcolsep) * \real{0.45}}\raggedright
\textbf{Parameter}\strut
\end{minipage} & \begin{minipage}[b]{(\columnwidth - 3\tabcolsep) * \real{0.16}}\centering
\textbf{R name}\strut
\end{minipage} & \begin{minipage}[b]{(\columnwidth - 3\tabcolsep) * \real{0.19}}\centering
\textbf{Base-case}\strut
\end{minipage} & \begin{minipage}[b]{(\columnwidth - 3\tabcolsep) * \real{0.20}}\centering
\textbf{Distribution}\strut
\end{minipage}\tabularnewline
\midrule
\endfirsthead
\toprule
\begin{minipage}[b]{(\columnwidth - 3\tabcolsep) * \real{0.45}}\raggedright
\textbf{Parameter}\strut
\end{minipage} & \begin{minipage}[b]{(\columnwidth - 3\tabcolsep) * \real{0.16}}\centering
\textbf{R name}\strut
\end{minipage} & \begin{minipage}[b]{(\columnwidth - 3\tabcolsep) * \real{0.19}}\centering
\textbf{Base-case}\strut
\end{minipage} & \begin{minipage}[b]{(\columnwidth - 3\tabcolsep) * \real{0.20}}\centering
\textbf{Distribution}\strut
\end{minipage}\tabularnewline
\midrule
\endhead
\begin{minipage}[t]{(\columnwidth - 3\tabcolsep) * \real{0.45}}\raggedright
Number of cycles (\(n_{T}\))\strut
\end{minipage} & \begin{minipage}[t]{(\columnwidth - 3\tabcolsep) * \real{0.16}}\centering
\texttt{n\_cycles}\strut
\end{minipage} & \begin{minipage}[t]{(\columnwidth - 3\tabcolsep) * \real{0.19}}\centering
75 years\strut
\end{minipage} & \begin{minipage}[t]{(\columnwidth - 3\tabcolsep) * \real{0.20}}\centering
-\strut
\end{minipage}\tabularnewline
\begin{minipage}[t]{(\columnwidth - 3\tabcolsep) * \real{0.45}}\raggedright
Names of health states (\(n\))\strut
\end{minipage} & \begin{minipage}[t]{(\columnwidth - 3\tabcolsep) * \real{0.16}}\centering
\texttt{v\_names\_states}\strut
\end{minipage} & \begin{minipage}[t]{(\columnwidth - 3\tabcolsep) * \real{0.19}}\centering
H, S1, S2, D\strut
\end{minipage} & \begin{minipage}[t]{(\columnwidth - 3\tabcolsep) * \real{0.20}}\centering
-\strut
\end{minipage}\tabularnewline
\begin{minipage}[t]{(\columnwidth - 3\tabcolsep) * \real{0.45}}\raggedright
Annual discount rate for costs\strut
\end{minipage} & \begin{minipage}[t]{(\columnwidth - 3\tabcolsep) * \real{0.16}}\centering
\texttt{d\_c}\strut
\end{minipage} & \begin{minipage}[t]{(\columnwidth - 3\tabcolsep) * \real{0.19}}\centering
3\%\strut
\end{minipage} & \begin{minipage}[t]{(\columnwidth - 3\tabcolsep) * \real{0.20}}\centering
-\strut
\end{minipage}\tabularnewline
\begin{minipage}[t]{(\columnwidth - 3\tabcolsep) * \real{0.45}}\raggedright
Annual discount rate for QALYs\strut
\end{minipage} & \begin{minipage}[t]{(\columnwidth - 3\tabcolsep) * \real{0.16}}\centering
\texttt{d\_e}\strut
\end{minipage} & \begin{minipage}[t]{(\columnwidth - 3\tabcolsep) * \real{0.19}}\centering
3\%\strut
\end{minipage} & \begin{minipage}[t]{(\columnwidth - 3\tabcolsep) * \real{0.20}}\centering
-\strut
\end{minipage}\tabularnewline
\begin{minipage}[t]{(\columnwidth - 3\tabcolsep) * \real{0.45}}\raggedright
Number of PSA samples (\(K\))\strut
\end{minipage} & \begin{minipage}[t]{(\columnwidth - 3\tabcolsep) * \real{0.16}}\centering
\texttt{n\_sim}\strut
\end{minipage} & \begin{minipage}[t]{(\columnwidth - 3\tabcolsep) * \real{0.19}}\centering
1,000\strut
\end{minipage} & \begin{minipage}[t]{(\columnwidth - 3\tabcolsep) * \real{0.20}}\centering
-\strut
\end{minipage}\tabularnewline
\begin{minipage}[t]{(\columnwidth - 3\tabcolsep) * \real{0.45}}\raggedright
Annual constant transition rates \strut
\end{minipage} & \begin{minipage}[t]{(\columnwidth - 3\tabcolsep) * \real{0.16}}\centering
\strut
\end{minipage} & \begin{minipage}[t]{(\columnwidth - 3\tabcolsep) * \real{0.19}}\centering
\strut
\end{minipage} & \begin{minipage}[t]{(\columnwidth - 3\tabcolsep) * \real{0.20}}\centering
\strut
\end{minipage}\tabularnewline
\begin{minipage}[t]{(\columnwidth - 3\tabcolsep) * \real{0.45}}\raggedright
- Disease onset (H to S1)\strut
\end{minipage} & \begin{minipage}[t]{(\columnwidth - 3\tabcolsep) * \real{0.16}}\centering
\texttt{r\_HS1}\strut
\end{minipage} & \begin{minipage}[t]{(\columnwidth - 3\tabcolsep) * \real{0.19}}\centering
0.150\strut
\end{minipage} & \begin{minipage}[t]{(\columnwidth - 3\tabcolsep) * \real{0.20}}\centering
gamma(30, 200)\strut
\end{minipage}\tabularnewline
\begin{minipage}[t]{(\columnwidth - 3\tabcolsep) * \real{0.45}}\raggedright
- Recovery (S1 to H)\strut
\end{minipage} & \begin{minipage}[t]{(\columnwidth - 3\tabcolsep) * \real{0.16}}\centering
\texttt{r\_S1H}\strut
\end{minipage} & \begin{minipage}[t]{(\columnwidth - 3\tabcolsep) * \real{0.19}}\centering
0.500\strut
\end{minipage} & \begin{minipage}[t]{(\columnwidth - 3\tabcolsep) * \real{0.20}}\centering
gamma(60, 120)\strut
\end{minipage}\tabularnewline
\begin{minipage}[t]{(\columnwidth - 3\tabcolsep) * \real{0.45}}\raggedright
- Disease progression (S1 to S2)\strut
\end{minipage} & \begin{minipage}[t]{(\columnwidth - 3\tabcolsep) * \real{0.16}}\centering
\texttt{r\_S1S2}\strut
\end{minipage} & \begin{minipage}[t]{(\columnwidth - 3\tabcolsep) * \real{0.19}}\centering
0.105\strut
\end{minipage} & \begin{minipage}[t]{(\columnwidth - 3\tabcolsep) * \real{0.20}}\centering
gamma(84, 800)\strut
\end{minipage}\tabularnewline
\begin{minipage}[t]{(\columnwidth - 3\tabcolsep) * \real{0.45}}\raggedright
Annual mortality\strut
\end{minipage} & \begin{minipage}[t]{(\columnwidth - 3\tabcolsep) * \real{0.16}}\centering
\strut
\end{minipage} & \begin{minipage}[t]{(\columnwidth - 3\tabcolsep) * \real{0.19}}\centering
\strut
\end{minipage} & \begin{minipage}[t]{(\columnwidth - 3\tabcolsep) * \real{0.20}}\centering
\strut
\end{minipage}\tabularnewline
\begin{minipage}[t]{(\columnwidth - 3\tabcolsep) * \real{0.45}}\raggedright
- Background mortality rate (H to D)\strut
\end{minipage} & \begin{minipage}[t]{(\columnwidth - 3\tabcolsep) * \real{0.16}}\centering
\texttt{r\_HD}\strut
\end{minipage} & \begin{minipage}[t]{(\columnwidth - 3\tabcolsep) * \real{0.19}}\centering
0.002\strut
\end{minipage} & \begin{minipage}[t]{(\columnwidth - 3\tabcolsep) * \real{0.20}}\centering
gamma(20, 10000)\strut
\end{minipage}\tabularnewline
\begin{minipage}[t]{(\columnwidth - 3\tabcolsep) * \real{0.45}}\raggedright
- Hazard ratio of death in S1 vs H\strut
\end{minipage} & \begin{minipage}[t]{(\columnwidth - 3\tabcolsep) * \real{0.16}}\centering
\texttt{hr\_S1}\strut
\end{minipage} & \begin{minipage}[t]{(\columnwidth - 3\tabcolsep) * \real{0.19}}\centering
3.0\strut
\end{minipage} & \begin{minipage}[t]{(\columnwidth - 3\tabcolsep) * \real{0.20}}\centering
lognormal(log(3.0), 0.01)\strut
\end{minipage}\tabularnewline
\begin{minipage}[t]{(\columnwidth - 3\tabcolsep) * \real{0.45}}\raggedright
- Hazard ratio of death in S2 vs H\strut
\end{minipage} & \begin{minipage}[t]{(\columnwidth - 3\tabcolsep) * \real{0.16}}\centering
\texttt{hr\_S2}\strut
\end{minipage} & \begin{minipage}[t]{(\columnwidth - 3\tabcolsep) * \real{0.19}}\centering
10.0\strut
\end{minipage} & \begin{minipage}[t]{(\columnwidth - 3\tabcolsep) * \real{0.20}}\centering
lognormal(log(10.0), 0.02)\strut
\end{minipage}\tabularnewline
\begin{minipage}[t]{(\columnwidth - 3\tabcolsep) * \real{0.45}}\raggedright
Annual costs\strut
\end{minipage} & \begin{minipage}[t]{(\columnwidth - 3\tabcolsep) * \real{0.16}}\centering
\strut
\end{minipage} & \begin{minipage}[t]{(\columnwidth - 3\tabcolsep) * \real{0.19}}\centering
\strut
\end{minipage} & \begin{minipage}[t]{(\columnwidth - 3\tabcolsep) * \real{0.20}}\centering
\strut
\end{minipage}\tabularnewline
\begin{minipage}[t]{(\columnwidth - 3\tabcolsep) * \real{0.45}}\raggedright
- Healthy individuals\strut
\end{minipage} & \begin{minipage}[t]{(\columnwidth - 3\tabcolsep) * \real{0.16}}\centering
\texttt{c\_H}\strut
\end{minipage} & \begin{minipage}[t]{(\columnwidth - 3\tabcolsep) * \real{0.19}}\centering
\$2,000\strut
\end{minipage} & \begin{minipage}[t]{(\columnwidth - 3\tabcolsep) * \real{0.20}}\centering
gamma(100.0, 20.0)\strut
\end{minipage}\tabularnewline
\begin{minipage}[t]{(\columnwidth - 3\tabcolsep) * \real{0.45}}\raggedright
- Sick individuals in S1\strut
\end{minipage} & \begin{minipage}[t]{(\columnwidth - 3\tabcolsep) * \real{0.16}}\centering
\texttt{c\_S1}\strut
\end{minipage} & \begin{minipage}[t]{(\columnwidth - 3\tabcolsep) * \real{0.19}}\centering
\$4,000\strut
\end{minipage} & \begin{minipage}[t]{(\columnwidth - 3\tabcolsep) * \real{0.20}}\centering
gamma(177.8, 22.5)\strut
\end{minipage}\tabularnewline
\begin{minipage}[t]{(\columnwidth - 3\tabcolsep) * \real{0.45}}\raggedright
- Sick individuals in S2\strut
\end{minipage} & \begin{minipage}[t]{(\columnwidth - 3\tabcolsep) * \real{0.16}}\centering
\texttt{c\_S2}\strut
\end{minipage} & \begin{minipage}[t]{(\columnwidth - 3\tabcolsep) * \real{0.19}}\centering
\$15,000\strut
\end{minipage} & \begin{minipage}[t]{(\columnwidth - 3\tabcolsep) * \real{0.20}}\centering
gamma(225.0, 66.7)\strut
\end{minipage}\tabularnewline
\begin{minipage}[t]{(\columnwidth - 3\tabcolsep) * \real{0.45}}\raggedright
- Dead individuals\strut
\end{minipage} & \begin{minipage}[t]{(\columnwidth - 3\tabcolsep) * \real{0.16}}\centering
\texttt{c\_D}\strut
\end{minipage} & \begin{minipage}[t]{(\columnwidth - 3\tabcolsep) * \real{0.19}}\centering
\$0\strut
\end{minipage} & \begin{minipage}[t]{(\columnwidth - 3\tabcolsep) * \real{0.20}}\centering
-\strut
\end{minipage}\tabularnewline
\begin{minipage}[t]{(\columnwidth - 3\tabcolsep) * \real{0.45}}\raggedright
Utility weights\strut
\end{minipage} & \begin{minipage}[t]{(\columnwidth - 3\tabcolsep) * \real{0.16}}\centering
\strut
\end{minipage} & \begin{minipage}[t]{(\columnwidth - 3\tabcolsep) * \real{0.19}}\centering
\strut
\end{minipage} & \begin{minipage}[t]{(\columnwidth - 3\tabcolsep) * \real{0.20}}\centering
\strut
\end{minipage}\tabularnewline
\begin{minipage}[t]{(\columnwidth - 3\tabcolsep) * \real{0.45}}\raggedright
- Healthy individuals\strut
\end{minipage} & \begin{minipage}[t]{(\columnwidth - 3\tabcolsep) * \real{0.16}}\centering
\texttt{u\_H}\strut
\end{minipage} & \begin{minipage}[t]{(\columnwidth - 3\tabcolsep) * \real{0.19}}\centering
1.00\strut
\end{minipage} & \begin{minipage}[t]{(\columnwidth - 3\tabcolsep) * \real{0.20}}\centering
beta(200, 3)\strut
\end{minipage}\tabularnewline
\begin{minipage}[t]{(\columnwidth - 3\tabcolsep) * \real{0.45}}\raggedright
- Sick individuals in S1\strut
\end{minipage} & \begin{minipage}[t]{(\columnwidth - 3\tabcolsep) * \real{0.16}}\centering
\texttt{u\_S1}\strut
\end{minipage} & \begin{minipage}[t]{(\columnwidth - 3\tabcolsep) * \real{0.19}}\centering
0.75\strut
\end{minipage} & \begin{minipage}[t]{(\columnwidth - 3\tabcolsep) * \real{0.20}}\centering
beta(130, 45)\strut
\end{minipage}\tabularnewline
\begin{minipage}[t]{(\columnwidth - 3\tabcolsep) * \real{0.45}}\raggedright
- Sick individuals in S2\strut
\end{minipage} & \begin{minipage}[t]{(\columnwidth - 3\tabcolsep) * \real{0.16}}\centering
\texttt{u\_S2}\strut
\end{minipage} & \begin{minipage}[t]{(\columnwidth - 3\tabcolsep) * \real{0.19}}\centering
0.50\strut
\end{minipage} & \begin{minipage}[t]{(\columnwidth - 3\tabcolsep) * \real{0.20}}\centering
beta(230, 230)\strut
\end{minipage}\tabularnewline
\begin{minipage}[t]{(\columnwidth - 3\tabcolsep) * \real{0.45}}\raggedright
- Dead individuals\strut
\end{minipage} & \begin{minipage}[t]{(\columnwidth - 3\tabcolsep) * \real{0.16}}\centering
\texttt{u\_D}\strut
\end{minipage} & \begin{minipage}[t]{(\columnwidth - 3\tabcolsep) * \real{0.19}}\centering
0.00\strut
\end{minipage} & \begin{minipage}[t]{(\columnwidth - 3\tabcolsep) * \real{0.20}}\centering
-\strut
\end{minipage}\tabularnewline
\begin{minipage}[t]{(\columnwidth - 3\tabcolsep) * \real{0.45}}\raggedright
Treatment A cost and effectiveness\strut
\end{minipage} & \begin{minipage}[t]{(\columnwidth - 3\tabcolsep) * \real{0.16}}\centering
\strut
\end{minipage} & \begin{minipage}[t]{(\columnwidth - 3\tabcolsep) * \real{0.19}}\centering
\strut
\end{minipage} & \begin{minipage}[t]{(\columnwidth - 3\tabcolsep) * \real{0.20}}\centering
\strut
\end{minipage}\tabularnewline
\begin{minipage}[t]{(\columnwidth - 3\tabcolsep) * \real{0.45}}\raggedright
- Cost of treatment A, additional to state-specific health care costs\strut
\end{minipage} & \begin{minipage}[t]{(\columnwidth - 3\tabcolsep) * \real{0.16}}\centering
\texttt{c\_trtA}\strut
\end{minipage} & \begin{minipage}[t]{(\columnwidth - 3\tabcolsep) * \real{0.19}}\centering
\$12,000\strut
\end{minipage} & \begin{minipage}[t]{(\columnwidth - 3\tabcolsep) * \real{0.20}}\centering
gamma(576.0, 20.8)\strut
\end{minipage}\tabularnewline
\begin{minipage}[t]{(\columnwidth - 3\tabcolsep) * \real{0.45}}\raggedright
- Utility for treated individuals in S1\strut
\end{minipage} & \begin{minipage}[t]{(\columnwidth - 3\tabcolsep) * \real{0.16}}\centering
\texttt{u\_trtA}\strut
\end{minipage} & \begin{minipage}[t]{(\columnwidth - 3\tabcolsep) * \real{0.19}}\centering
0.95\strut
\end{minipage} & \begin{minipage}[t]{(\columnwidth - 3\tabcolsep) * \real{0.20}}\centering
beta(300, 15)\strut
\end{minipage}\tabularnewline
\begin{minipage}[t]{(\columnwidth - 3\tabcolsep) * \real{0.45}}\raggedright
Treatment B cost and effectiveness\strut
\end{minipage} & \begin{minipage}[t]{(\columnwidth - 3\tabcolsep) * \real{0.16}}\centering
\strut
\end{minipage} & \begin{minipage}[t]{(\columnwidth - 3\tabcolsep) * \real{0.19}}\centering
\strut
\end{minipage} & \begin{minipage}[t]{(\columnwidth - 3\tabcolsep) * \real{0.20}}\centering
\strut
\end{minipage}\tabularnewline
\begin{minipage}[t]{(\columnwidth - 3\tabcolsep) * \real{0.45}}\raggedright
- Cost of treatment B, additional to state-specific health care costs\strut
\end{minipage} & \begin{minipage}[t]{(\columnwidth - 3\tabcolsep) * \real{0.16}}\centering
\texttt{c\_trtB}\strut
\end{minipage} & \begin{minipage}[t]{(\columnwidth - 3\tabcolsep) * \real{0.19}}\centering
\$12,000\strut
\end{minipage} & \begin{minipage}[t]{(\columnwidth - 3\tabcolsep) * \real{0.20}}\centering
gamma(676.0, 19.2)\strut
\end{minipage}\tabularnewline
\begin{minipage}[t]{(\columnwidth - 3\tabcolsep) * \real{0.45}}\raggedright
- Reduction in rate of disease progression (S1 to S2) as hazard ratio (HR)\strut
\end{minipage} & \begin{minipage}[t]{(\columnwidth - 3\tabcolsep) * \real{0.16}}\centering
\texttt{hr\_S1S2\_trtB}\strut
\end{minipage} & \begin{minipage}[t]{(\columnwidth - 3\tabcolsep) * \real{0.19}}\centering
log(0.6)\strut
\end{minipage} & \begin{minipage}[t]{(\columnwidth - 3\tabcolsep) * \real{0.20}}\centering
lognormal(log(0.6), 0.1)\strut
\end{minipage}\tabularnewline
\bottomrule
\end{longtable}

The following sections include R code snippets. All R code is stored as a GitHub repository and can be accessed at  \url{https://github.com/DARTH-git/cohort-modeling-tutorial-intro}. We initialize the input parameters in the R code below by setting the variables to their base-case values. We do this process as the first coding step, all in one place, so the updated value will carry through the rest of the code when a parameter value changes.

\begin{Shaded}
\begin{Highlighting}[]
\DocumentationTok{\#\# General setup}
\NormalTok{cycle\_length }\OtherTok{\textless{}{-}} \DecValTok{1} \CommentTok{\# cycle length equal one year (use 1/12 for monthly)}
\NormalTok{n\_age\_init }\OtherTok{\textless{}{-}} \DecValTok{25}  \CommentTok{\# age at baseline}
\NormalTok{n\_age\_max  }\OtherTok{\textless{}{-}} \DecValTok{100} \CommentTok{\# maximum age of follow up}
\NormalTok{n\_cycles }\OtherTok{\textless{}{-}}\NormalTok{ (n\_age\_max }\SpecialCharTok{{-}}\NormalTok{ n\_age\_init)}\SpecialCharTok{/}\NormalTok{cycle\_length }\CommentTok{\# time horizon, number of cycles}
\NormalTok{v\_names\_states }\OtherTok{\textless{}{-}} \FunctionTok{c}\NormalTok{(}\StringTok{"H"}\NormalTok{, }\StringTok{"S1"}\NormalTok{, }\StringTok{"S2"}\NormalTok{, }\StringTok{"D"}\NormalTok{) }\CommentTok{\# the 4 health states of the model:}
                               \CommentTok{\# Healthy (H), Sick (S1), Sicker (S2), Dead (D)}
\NormalTok{n\_states }\OtherTok{\textless{}{-}} \FunctionTok{length}\NormalTok{(v\_names\_states) }\CommentTok{\# number of health states }
\NormalTok{d\_e }\OtherTok{\textless{}{-}} \FloatTok{0.03} \CommentTok{\# annual discount rate for QALYs of 3\% }
\NormalTok{d\_c }\OtherTok{\textless{}{-}} \FloatTok{0.03} \CommentTok{\# annual discount rate for costs of 3\% }
\NormalTok{v\_names\_str }\OtherTok{\textless{}{-}} \FunctionTok{c}\NormalTok{(}\StringTok{"Standard of care"}\NormalTok{, }\CommentTok{\# store the strategy names}
                 \StringTok{"Strategy A"}\NormalTok{, }
                 \StringTok{"Strategy B"}\NormalTok{,}
                 \StringTok{"Strategy AB"}\NormalTok{) }

\DocumentationTok{\#\# Transition probabilities (annual), and hazard ratios (HRs)}
\NormalTok{r\_HD   }\OtherTok{\textless{}{-}} \FloatTok{0.002} \CommentTok{\# constant annual rate of dying when Healthy (all{-}cause mortality rate)}
\NormalTok{r\_HS1  }\OtherTok{\textless{}{-}} \FloatTok{0.15}  \CommentTok{\# constant annual rate of becoming Sick when Healthy}
\NormalTok{r\_S1H  }\OtherTok{\textless{}{-}} \FloatTok{0.5}   \CommentTok{\# constant annual rate of becoming Healthy when Sick}
\NormalTok{r\_S1S2 }\OtherTok{\textless{}{-}} \FloatTok{0.105} \CommentTok{\# constant annual rate of becoming Sicker when Sick}
\NormalTok{hr\_S1  }\OtherTok{\textless{}{-}} \DecValTok{3}     \CommentTok{\# hazard ratio of death in Sick vs Healthy}
\NormalTok{hr\_S2  }\OtherTok{\textless{}{-}} \DecValTok{10}    \CommentTok{\# hazard ratio of death in Sicker vs Healthy }

\DocumentationTok{\#\#\# Process model inputs}
\DocumentationTok{\#\# Constant transition probability of becoming Sick when Healthy }
\CommentTok{\# transform rate to probability and scale by the cycle length}
\NormalTok{p\_HS1 }\OtherTok{\textless{}{-}} \DecValTok{1} \SpecialCharTok{{-}} \FunctionTok{exp}\NormalTok{(}\SpecialCharTok{{-}}\NormalTok{r\_HS1 }\SpecialCharTok{*}\NormalTok{ cycle\_length)}
\DocumentationTok{\#\# Constant transition probability of becoming Healthy when Sick }
\CommentTok{\# transform rate to probability and scale by the cycle length}
\NormalTok{p\_S1H }\OtherTok{\textless{}{-}} \DecValTok{1} \SpecialCharTok{{-}} \FunctionTok{exp}\NormalTok{(}\SpecialCharTok{{-}}\NormalTok{r\_S1H }\SpecialCharTok{*}\NormalTok{ cycle\_length)}
\DocumentationTok{\#\# Constant transition probability of becoming Sicker when Sick }
\CommentTok{\# transform rate to probability and scale by the cycle length}
\NormalTok{p\_S1S2 }\OtherTok{\textless{}{-}} \DecValTok{1} \SpecialCharTok{{-}} \FunctionTok{exp}\NormalTok{(}\SpecialCharTok{{-}}\NormalTok{r\_S1S2 }\SpecialCharTok{*}\NormalTok{ cycle\_length)}

\CommentTok{\# Effectiveness of treatment B}
\NormalTok{hr\_S1S2\_trtB }\OtherTok{\textless{}{-}} \FloatTok{0.6} \CommentTok{\# hazard ratio of becoming Sicker when Sick under treatment B}

\DocumentationTok{\#\# State rewards}
\DocumentationTok{\#\# Costs}
\NormalTok{c\_H    }\OtherTok{\textless{}{-}} \DecValTok{2000}  \CommentTok{\# annual cost of being Healthy}
\NormalTok{c\_S1   }\OtherTok{\textless{}{-}} \DecValTok{4000}  \CommentTok{\# annual cost of being Sick}
\NormalTok{c\_S2   }\OtherTok{\textless{}{-}} \DecValTok{15000} \CommentTok{\# annual cost of being Sicker}
\NormalTok{c\_D    }\OtherTok{\textless{}{-}} \DecValTok{0}     \CommentTok{\# annual cost of being dead}
\NormalTok{c\_trtA }\OtherTok{\textless{}{-}} \DecValTok{12000} \CommentTok{\# annual cost of receiving treatment A}
\NormalTok{c\_trtB }\OtherTok{\textless{}{-}} \DecValTok{13000} \CommentTok{\# annual cost of receiving treatment B}
\CommentTok{\# Utilities}
\NormalTok{u\_H    }\OtherTok{\textless{}{-}} \DecValTok{1}    \CommentTok{\# annual utility of being Healthy}
\NormalTok{u\_S1   }\OtherTok{\textless{}{-}} \FloatTok{0.75} \CommentTok{\# annual utility of being Sick}
\NormalTok{u\_S2   }\OtherTok{\textless{}{-}} \FloatTok{0.5}  \CommentTok{\# annual utility of being Sicker}
\NormalTok{u\_D    }\OtherTok{\textless{}{-}} \DecValTok{0}    \CommentTok{\# annual utility of being dead}
\NormalTok{u\_trtA }\OtherTok{\textless{}{-}} \FloatTok{0.95} \CommentTok{\# annual utility when receiving treatment A}
\end{Highlighting}
\end{Shaded}

To compute the background mortality risk, \texttt{p\_HD}, from the background mortality rate for the same cycle length (i.e., \texttt{cycle\_length = 1}), we apply Equation \eqref{eq:rate-to-prob-ann} to \texttt{r\_HD}. To compute the mortality risks of the cohort in S1 and S2, we multiply the background mortality rate \texttt{r\_HD} by the hazard ratios \texttt{hr\_S1} and \texttt{hr\_S2}, respectively, and then convert back to probabilities using Equation \eqref{eq:rate-to-prob-ann}. These calculations are required because hazard ratios only apply to rates and not to probabilities. The code below performs the computation in R. In the \texttt{darthtools} package (\url{https://github.com/DARTH-git/darthtools}), we provide R functions that compute transformations between rates and probabilities since these transformations are frequently used.

\begin{Shaded}
\begin{Highlighting}[]
\DocumentationTok{\#\# Mortality rates}
\NormalTok{r\_S1D }\OtherTok{\textless{}{-}}\NormalTok{ r\_HD }\SpecialCharTok{*}\NormalTok{ hr\_S1 }\CommentTok{\# annual rate of dying when Sick}
\NormalTok{r\_S2D }\OtherTok{\textless{}{-}}\NormalTok{ r\_HD }\SpecialCharTok{*}\NormalTok{ hr\_S2 }\CommentTok{\# annual rate of dying when Sicker}
\DocumentationTok{\#\# Cycle-specific probabilities of dying}
\NormalTok{cycle\_length }\OtherTok{\textless{}{-}} \DecValTok{1}
\NormalTok{p\_HD  }\OtherTok{\textless{}{-}} \DecValTok{1} \SpecialCharTok{{-}} \FunctionTok{exp}\NormalTok{(}\SpecialCharTok{{-}}\NormalTok{r\_HD}\SpecialCharTok{ * }\NormalTok{cycle\_length)  }\CommentTok{\# annual background mortality risk (i.e., probability)}
\NormalTok{p\_S1D }\OtherTok{\textless{}{-}} \DecValTok{1} \SpecialCharTok{{-}} \FunctionTok{exp}\NormalTok{(}\SpecialCharTok{{-}}\NormalTok{r\_S1D}\SpecialCharTok{ * }\NormalTok{cycle\_length) }\CommentTok{\# annual probability of dying when Sick}
\NormalTok{p\_S2D }\OtherTok{\textless{}{-}} \DecValTok{1} \SpecialCharTok{{-}} \FunctionTok{exp}\NormalTok{(}\SpecialCharTok{{-}}\NormalTok{r\_S2D}\SpecialCharTok{ * }\NormalTok{cycle\_length) }\CommentTok{\# annual probability of dying when Sicker}
\end{Highlighting}
\end{Shaded}

To compute the risk of progression from S1 to S2 under treatment B, we multiply the hazard ratio of treatment B by the rate of progressing from S1 to S2 and transform it to probability by applying Equation \eqref{eq:rate-to-prob-ann}.

\begin{Shaded}
\begin{Highlighting}[]
\DocumentationTok{\#\# Transition probability of becoming Sicker when Sick for treatment B}
\CommentTok{\# apply hazard ratio to rate to obtain transition rate of becoming Sicker when Sick }
\CommentTok{\# for treatment B}
\NormalTok{r\_S1S2\_trtB }\OtherTok{\textless{}{-}}\NormalTok{ r\_S1S2 }\SpecialCharTok{*}\NormalTok{ hr\_S1S2\_trtB}
\CommentTok{\# transform rate to probability}
\CommentTok{\# probability to become Sicker when Sick under treatment B  }
\NormalTok{p\_S1S2\_trtB }\OtherTok{\textless{}{-}} \DecValTok{1} \SpecialCharTok{{-}} \FunctionTok{exp}\NormalTok{(}\SpecialCharTok{{-}}\NormalTok{r\_S1S2\_trtB }\SpecialCharTok{*}\NormalTok{ cycle\_length) }
\end{Highlighting}
\end{Shaded}

For the Sick-Sicker model, the entire cohort starts in the H state. Therefore, we create the \(1 \times n_S\) initial state vector \texttt{v\_m\_init} with all of the cohort assigned to the H state:

\begin{Shaded}
\begin{Highlighting}[]
\NormalTok{v\_m\_init }\OtherTok{\textless{}{-}} \FunctionTok{c}\NormalTok{(}\AttributeTok{H =} \DecValTok{1}\NormalTok{, }\AttributeTok{S1 =} \DecValTok{0}\NormalTok{, }\AttributeTok{S2 =} \DecValTok{0}\NormalTok{, }\AttributeTok{D =} \DecValTok{0}\NormalTok{) }\CommentTok{\# initial state vector}
\end{Highlighting}
\end{Shaded}

The variable \texttt{v\_m\_init} is used to initialize \(M\) represented by \texttt{m\_M} for the cohort under strategy SoC. We also create a trace for each of the other treatment-based strategies.

\begin{Shaded}
\begin{Highlighting}[]
\DocumentationTok{\#\# Initialize cohort trace for SoC}
\NormalTok{m\_M }\OtherTok{\textless{}{-}} \FunctionTok{matrix}\NormalTok{(}\ConstantTok{NA}\NormalTok{, }
              \AttributeTok{nrow =}\NormalTok{ (n\_cycles }\SpecialCharTok{+} \DecValTok{1}\NormalTok{), }\AttributeTok{ncol =}\NormalTok{ n\_states, }
              \AttributeTok{dimnames =} \FunctionTok{list}\NormalTok{(}\DecValTok{0}\SpecialCharTok{:}\NormalTok{n\_cycles, v\_names\_states))}
\CommentTok{\# Store the initial state vector in the first row of the cohort trace}
\NormalTok{m\_M[}\DecValTok{1}\NormalTok{, ] }\OtherTok{\textless{}{-}}\NormalTok{ v\_m\_init}
\DocumentationTok{\#\# Initialize cohort trace for strategies A, B, and AB}
\CommentTok{\# Structure and initial states are the same as for SoC}
\NormalTok{m\_M\_strA  }\OtherTok{\textless{}{-}}\NormalTok{ m\_M }\CommentTok{\# Strategy A}
\NormalTok{m\_M\_strB  }\OtherTok{\textless{}{-}}\NormalTok{ m\_M }\CommentTok{\# Strategy B}
\NormalTok{m\_M\_strAB }\OtherTok{\textless{}{-}}\NormalTok{ m\_M }\CommentTok{\# Strategy AB}
\end{Highlighting}
\end{Shaded}

Note that the initial state vector, \texttt{v\_m\_init}, can be modified to account for the cohort's distribution across the states at the start of the simulation. This distribution can also vary by strategy if needed.

Since the Sick-Sicker model consists of 4 states, we create a 4 \(\times\) 4 transition probability matrix for strategy SoC, \texttt{m\_P}. We initialize the matrix with default values of zero for all transition probabilities and then populate it with the corresponding transition probabilities. To access an element of \texttt{m\_P}, we specify first the row name (or number) and then the column name (or number) separated by a comma. For example, we could access the transition probability from state Healthy (H) to state Sick (S1) using the corresponding row or column state-names as characters \texttt{m\_P{[}"H",\ "S1"{]}}. We assume that all transitions to non-death states are conditional on surviving to the end of a cycle. Thus, we first condition on surviving by multiplying the transition probabilities times \texttt{1\ -\ p\_HD}, the probability of surviving a cycle. For example, to obtain the probability of transitioning from H to S1, we multiply the transition probability from H to S1 conditional on being alive, \texttt{p\_HS1} by \texttt{1\ -\ p\_HD}. We create the transition probability matrix for strategy A as a copy of the SoC's transition probability matrix because treatment A does not alter the cohort's transition probabilities.

\begin{Shaded}
\begin{Highlighting}[]
\DocumentationTok{\#\# Initialize transition probability matrix for strategy SoC}
\NormalTok{m\_P }\OtherTok{\textless{}{-}} \FunctionTok{matrix}\NormalTok{(}\DecValTok{0}\NormalTok{, }
              \AttributeTok{nrow =}\NormalTok{ n\_states, }\AttributeTok{ncol =}\NormalTok{ n\_states, }
              \AttributeTok{dimnames =} \FunctionTok{list}\NormalTok{(v\_names\_states, v\_names\_states)) }\CommentTok{\# row and column names}
\DocumentationTok{\#\# Fill in matrix}
\CommentTok{\# From H}
\NormalTok{m\_P[}\StringTok{"H"}\NormalTok{, }\StringTok{"H"}\NormalTok{]   }\OtherTok{\textless{}{-}}\NormalTok{ (}\DecValTok{1} \SpecialCharTok{{-}}\NormalTok{ p\_HD) }\SpecialCharTok{*}\NormalTok{ (}\DecValTok{1} \SpecialCharTok{{-}}\NormalTok{ p\_HS1)}
\NormalTok{m\_P[}\StringTok{"H"}\NormalTok{, }\StringTok{"S1"}\NormalTok{]  }\OtherTok{\textless{}{-}}\NormalTok{ (}\DecValTok{1} \SpecialCharTok{{-}}\NormalTok{ p\_HD) }\SpecialCharTok{*}\NormalTok{ p\_HS1}
\NormalTok{m\_P[}\StringTok{"H"}\NormalTok{, }\StringTok{"D"}\NormalTok{]   }\OtherTok{\textless{}{-}}\NormalTok{ p\_HD}
\CommentTok{\# From S1}
\NormalTok{m\_P[}\StringTok{"S1"}\NormalTok{, }\StringTok{"H"}\NormalTok{]  }\OtherTok{\textless{}{-}}\NormalTok{ (}\DecValTok{1} \SpecialCharTok{{-}}\NormalTok{ p\_S1D) }\SpecialCharTok{*}\NormalTok{ p\_S1H}
\NormalTok{m\_P[}\StringTok{"S1"}\NormalTok{, }\StringTok{"S1"}\NormalTok{] }\OtherTok{\textless{}{-}}\NormalTok{ (}\DecValTok{1} \SpecialCharTok{{-}}\NormalTok{ p\_S1D) }\SpecialCharTok{*}\NormalTok{ (}\DecValTok{1} \SpecialCharTok{{-}}\NormalTok{ (p\_S1H }\SpecialCharTok{+}\NormalTok{ p\_S1S2))}
\NormalTok{m\_P[}\StringTok{"S1"}\NormalTok{, }\StringTok{"S2"}\NormalTok{] }\OtherTok{\textless{}{-}}\NormalTok{ (}\DecValTok{1} \SpecialCharTok{{-}}\NormalTok{ p\_S1D) }\SpecialCharTok{*}\NormalTok{ p\_S1S2}
\NormalTok{m\_P[}\StringTok{"S1"}\NormalTok{, }\StringTok{"D"}\NormalTok{]  }\OtherTok{\textless{}{-}}\NormalTok{ p\_S1D}
\CommentTok{\# From S2}
\NormalTok{m\_P[}\StringTok{"S2"}\NormalTok{, }\StringTok{"S2"}\NormalTok{] }\OtherTok{\textless{}{-}} \DecValTok{1} \SpecialCharTok{{-}}\NormalTok{ p\_S2D}
\NormalTok{m\_P[}\StringTok{"S2"}\NormalTok{, }\StringTok{"D"}\NormalTok{]  }\OtherTok{\textless{}{-}}\NormalTok{ p\_S2D}
\CommentTok{\# From D}
\NormalTok{m\_P[}\StringTok{"D"}\NormalTok{, }\StringTok{"D"}\NormalTok{]   }\OtherTok{\textless{}{-}} \DecValTok{1}

\DocumentationTok{\#\# Initialize transition probability matrix for strategy A as a copy of SoC\textquotesingle{}s}
\NormalTok{m\_P\_strA }\OtherTok{\textless{}{-}}\NormalTok{ m\_P}
\end{Highlighting}
\end{Shaded}

Because treatment B alters progression from S1 to S2, we created a different transition probability matrix to model this treatment, \texttt{m\_P\_strB}. We initialize \texttt{m\_P\_strB} as a copy of \texttt{m\_P} and update only the transition probabilities from S1 to S2 (i.e., \texttt{p\_S1S2} is replaced with \texttt{p\_S1S2\_trtB}). Strategy AB also alters progression from S1 to S2 because it uses treatment B, so we create this strategy's transition probability matrix as a copy of the transition probability matrix of strategy B.

\begin{Shaded}
\begin{Highlighting}[]
\DocumentationTok{\#\# Initialize transition probability matrix for strategy B}
\NormalTok{m\_P\_strB }\OtherTok{\textless{}{-}}\NormalTok{ m\_P}
\DocumentationTok{\#\# Update only transition probabilities from S1 involving p\_S1S2}
\NormalTok{m\_P\_strB[}\StringTok{"S1"}\NormalTok{, }\StringTok{"S1"}\NormalTok{] }\OtherTok{\textless{}{-}}\NormalTok{ (}\DecValTok{1} \SpecialCharTok{{-}}\NormalTok{ p\_S1D) }\SpecialCharTok{*}\NormalTok{ (}\DecValTok{1} \SpecialCharTok{{-}}\NormalTok{ (p\_S1H }\SpecialCharTok{+}\NormalTok{ p\_S1S2\_trtB))}
\NormalTok{m\_P\_strB[}\StringTok{"S1"}\NormalTok{, }\StringTok{"S2"}\NormalTok{] }\OtherTok{\textless{}{-}}\NormalTok{ (}\DecValTok{1} \SpecialCharTok{{-}}\NormalTok{ p\_S1D) }\SpecialCharTok{*}\NormalTok{ p\_S1S2\_trtB}

\DocumentationTok{\#\# Initialize transition probability matrix for strategy AB as a copy of B\textquotesingle{}s}
\NormalTok{m\_P\_strAB }\OtherTok{\textless{}{-}}\NormalTok{ m\_P\_strB}
\end{Highlighting}
\end{Shaded}

Once all transition matrices are created, we verify they are valid by checking that each row sums to one and that each entry is between 0 and 1. In the \texttt{darthtools} package (\url{https://github.com/DARTH-git/darthtools}), we provide R functions that do these checks and have been described previously.\textsuperscript{\protect\hyperlink{ref-Alarid-Escudero2019e}{20}}

\begin{Shaded}
\begin{Highlighting}[]
\DocumentationTok{\#\#\# Check if transition probability matrices are valid}
\DocumentationTok{\#\# Check that transition probabilities are [0, 1]}
\NormalTok{m\_P }\SpecialCharTok{\textgreater{}=} \DecValTok{0} \SpecialCharTok{\&\&}\NormalTok{ m\_P }\SpecialCharTok{\textless{}=} \DecValTok{1}
\NormalTok{m\_P\_strA }\SpecialCharTok{\textgreater{}=} \DecValTok{0} \SpecialCharTok{\&\&}\NormalTok{ m\_P\_strA }\SpecialCharTok{\textless{}=} \DecValTok{1}
\NormalTok{m\_P\_strB }\SpecialCharTok{\textgreater{}=} \DecValTok{0} \SpecialCharTok{\&\&}\NormalTok{ m\_P\_strB }\SpecialCharTok{\textless{}=} \DecValTok{1}
\NormalTok{m\_P\_strAB }\SpecialCharTok{\textgreater{}=} \DecValTok{0} \SpecialCharTok{\&\&}\NormalTok{ m\_P\_strAB }\SpecialCharTok{\textless{}=} \DecValTok{1}
\DocumentationTok{\#\# Check that all rows sum to 1}
\FunctionTok{rowSums}\NormalTok{(m\_P) }\SpecialCharTok{==} \DecValTok{1}
\FunctionTok{rowSums}\NormalTok{(m\_P\_strA) }\SpecialCharTok{==} \DecValTok{1}
\FunctionTok{rowSums}\NormalTok{(m\_P\_strB) }\SpecialCharTok{==} \DecValTok{1}
\FunctionTok{rowSums}\NormalTok{(m\_P\_strAB) }\SpecialCharTok{==} \DecValTok{1}
\end{Highlighting}
\end{Shaded}

Next, we obtain the cohort distribution across the 4 states over 75 cycles using a time-independent cSTM under all four strategies. To achieve this, we iteratively compute the matrix product between each of the rows of \texttt{m\_M} and \texttt{m\_P}, and between \texttt{m\_M\_strB} and \texttt{m\_P\_strB}, respectively, using the \texttt{\%*\%} symbol in R at each cycle using a \texttt{for} loop

\begin{Shaded}
\begin{Highlighting}[]
\CommentTok{\# Iterative solution of time{-}independent cSTM}
\ControlFlowTok{for}\NormalTok{(t }\ControlFlowTok{in} \DecValTok{1}\SpecialCharTok{:}\NormalTok{n\_cycles)\{}
  \CommentTok{\# For SoC}
\NormalTok{  m\_M[t }\SpecialCharTok{+} \DecValTok{1}\NormalTok{, ] }\OtherTok{\textless{}{-}}\NormalTok{ m\_M[t, ] }\SpecialCharTok{\%*\%}\NormalTok{ m\_P}
  \CommentTok{\# For strategy A}
\NormalTok{  m\_M\_strA[t }\SpecialCharTok{+} \DecValTok{1}\NormalTok{, ] }\OtherTok{\textless{}{-}}\NormalTok{ m\_M\_strA[t, ] }\SpecialCharTok{\%*\%}\NormalTok{ m\_P\_strA}
  \CommentTok{\# For strategy B}
\NormalTok{  m\_M\_strB[t }\SpecialCharTok{+} \DecValTok{1}\NormalTok{, ] }\OtherTok{\textless{}{-}}\NormalTok{ m\_M\_strB[t, ] }\SpecialCharTok{\%*\%}\NormalTok{ m\_P\_strB}
  \CommentTok{\# For strategy AB}
\NormalTok{  m\_M\_strAB[t }\SpecialCharTok{+} \DecValTok{1}\NormalTok{, ] }\OtherTok{\textless{}{-}}\NormalTok{ m\_M\_strAB[t, ] }\SpecialCharTok{\%*\%}\NormalTok{ m\_P\_strAB}
\NormalTok{\}}
\end{Highlighting}
\end{Shaded}

\begin{figure}[H]

{\centering \includegraphics{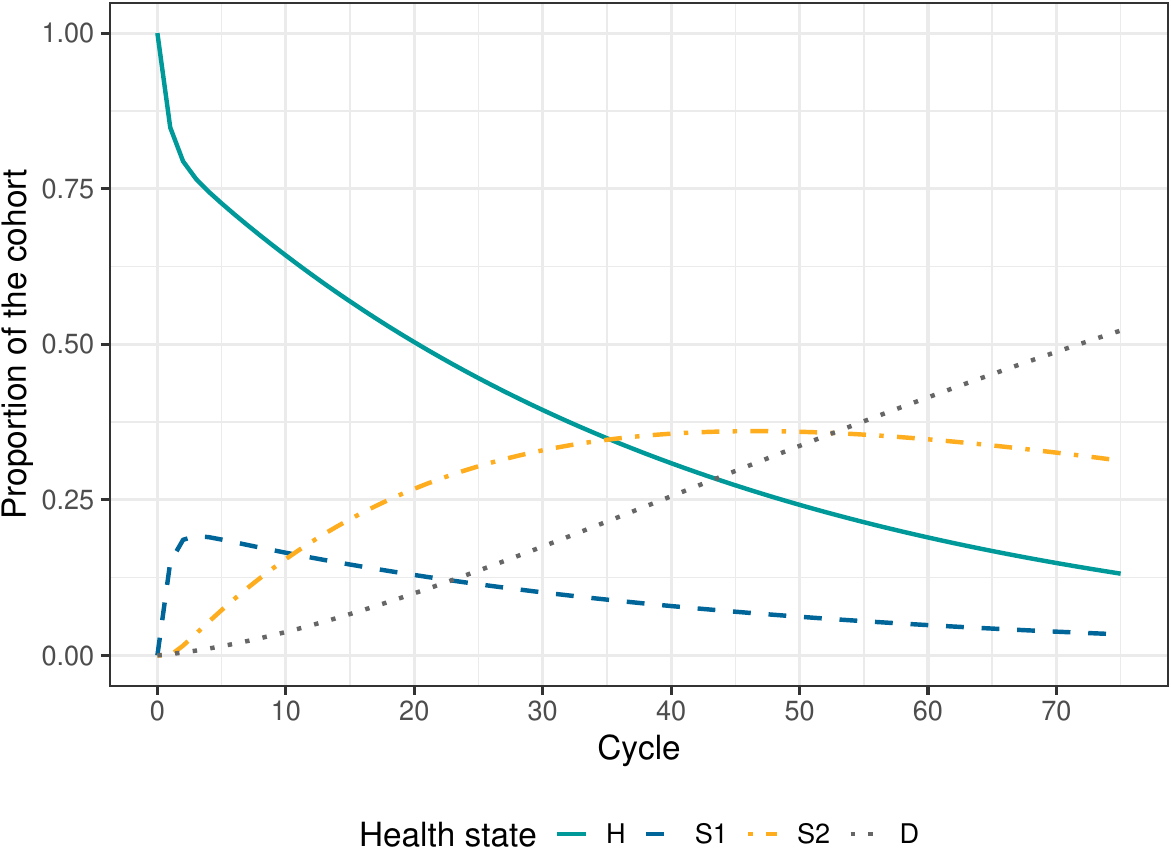} 

}

\caption{Cohort trace of the time-independent cSTM under strategies SoC and A.}\label{fig:Sick-Sicker-Trace-TimeHom}
\end{figure}

Table \ref{tab:Trace} shows the cohort trace matrix \(M\) of the Sick-Sicker model under strategies SoC and A for the first six cycles. The whole cohort starts in the H state and transitions to the rest of the states over time. Given that the D state is absorbing, the proportion in this state increases over time. A graphical representation of the cohort trace for all the cycles is shown in Figure \ref{fig:Sick-Sicker-Trace-TimeHom}.

\begin{table}[!h]

\caption{\label{tab:Trace}The distribution of the cohort under strategies SoC and A for the first six cycles of the time-independent Sick-Sicker model. The first row, labeled with cycle 0, contains the distribution of the cohort at time zero.}
\centering
\begin{tabular}[t]{ccccc}
\toprule{}
Cycle & H & S1 & S2 & D\\
\midrule{}
0 & 1.000 & 0.000 & 0.000 & 0.000\\
1 & 0.859 & 0.139 & 0.000 & 0.002\\
2 & 0.792 & 0.189 & 0.014 & 0.005\\
3 & 0.755 & 0.206 & 0.032 & 0.008\\
4 & 0.729 & 0.208 & 0.052 & 0.011\\
5 & 0.707 & 0.206 & 0.072 & 0.015\\
\bottomrule{}
\end{tabular}
\end{table}

\hypertarget{cost-and-effectiveness-outcomes}{%
\section{Cost and effectiveness outcomes}\label{cost-and-effectiveness-outcomes}}

We are interested in computing the total QALYs and costs accrued by the cohort over a predefined time horizon for a CEA. In the advanced cSTM tutorial,\textsuperscript{\protect\hyperlink{ref-Alarid-Escudero2021b}{16}} we describe how to compute epidemiological outcomes from cSTMs, such as survival, prevalence, and life expectancy.\textsuperscript{\protect\hyperlink{ref-Siebert2012c}{2}} These epidemiological outcomes are often used to produce other measures of interest for model calibration and validation.

\hypertarget{state-rewards}{%
\subsection{State rewards}\label{state-rewards}}

A state reward refers to a value assigned to individuals for being in a given health state. These could be either utilities or costs associated with remaining in a specific health state for one cycle in a CEA context. The column vector \(\mathbf{y}\) of size \(n_T+1\) can represent the total expected reward of an outcome of interest for the entire cohort at each cycle. To calculate \(\mathbf{y}\), we compute the matrix product of the cohort trace matrix times a \emph{vector} of state rewards \(\mathbf{r}\) of the same dimension as the number of states (\(n_S\)), such that
\begin{equation}
  \mathbf{y} = M\mathbf{r}.
  \label{eq:exp-rew-cycle}
\end{equation}

For the Sick-Sicker model, we create a vector of utilities and costs for each of the four strategies considered. The vectors of utilities and costs in R, \texttt{v\_u\_SoC} and \texttt{v\_c\_SoC}, respectively, represent the utilities and costs in each of the four health states under SoC, scaled by the cycle length (values are shown in Table \ref{tab:param-table}).

\begin{Shaded}
\begin{Highlighting}[]
\CommentTok{\# Vector of state utilities under SOC}
\NormalTok{v\_u\_SoC }\OtherTok{\textless{}{-}} \FunctionTok{c}\NormalTok{(}\AttributeTok{H =}\NormalTok{ u\_H, }\AttributeTok{S1 =}\NormalTok{ u\_S1, }\AttributeTok{S2 =}\NormalTok{ u\_S2, }\AttributeTok{D =}\NormalTok{ u\_D) }\SpecialCharTok{*}\NormalTok{ cycle\_length }
\CommentTok{\# Vector of state costs under SoC}
\NormalTok{v\_c\_SoC }\OtherTok{\textless{}{-}} \FunctionTok{c}\NormalTok{(}\AttributeTok{H =}\NormalTok{ c\_H, }\AttributeTok{S1 =}\NormalTok{ c\_S1, }\AttributeTok{S2 =}\NormalTok{ c\_S2, }\AttributeTok{D =}\NormalTok{ c\_D) }\SpecialCharTok{*}\NormalTok{ cycle\_length }
\end{Highlighting}
\end{Shaded}

We account for the benefits and costs of both treatments individually and their combination to create the state-reward vectors under treatments A and B (strategies A and B, respectively) and when applied jointly (strategy AB). Only treatment A affects QoL, so we create a vector of utilities specific to strategies involving treatment A (strategies A and AB), \texttt{v\_u\_strA} and \texttt{v\_u\_strAB}. These vectors will have the same utility weights as for strategy SoC except for being in S1. We assign the utility associated with the benefit of treatment A in that state, \texttt{u\_trtA}. Treatment B does not affect QoL, so the vector of utilities for strategy involving treatment B, \texttt{v\_u\_strB}, is the same as for SoC.

\begin{Shaded}
\begin{Highlighting}[]
\CommentTok{\# Vector of state utilities for strategy A}
\NormalTok{v\_u\_strA }\OtherTok{\textless{}{-}} \FunctionTok{c}\NormalTok{(}\AttributeTok{H =}\NormalTok{ u\_H, }\AttributeTok{S1 =}\NormalTok{ u\_trtA, }\AttributeTok{S2 =}\NormalTok{ u\_S2, }\AttributeTok{D =}\NormalTok{ u\_D) }\SpecialCharTok{*}\NormalTok{ cycle\_length }
\CommentTok{\# Vector of state utilities for strategy B}
\NormalTok{v\_u\_strB }\OtherTok{\textless{}{-}} \FunctionTok{c}\NormalTok{(}\AttributeTok{H =}\NormalTok{ u\_H, }\AttributeTok{S1 =}\NormalTok{ u\_S1, }\AttributeTok{S2 =}\NormalTok{ u\_S2, }\AttributeTok{D =}\NormalTok{ u\_D) }\SpecialCharTok{*}\NormalTok{ cycle\_length }
\CommentTok{\# Vector of state utilities for strategy AB}
\NormalTok{v\_u\_strAB }\OtherTok{\textless{}{-}} \FunctionTok{c}\NormalTok{(}\AttributeTok{H =}\NormalTok{ u\_H, }\AttributeTok{S1 =}\NormalTok{ u\_trtA, }\AttributeTok{S2 =}\NormalTok{ u\_S2, }\AttributeTok{D =}\NormalTok{ u\_D) }\SpecialCharTok{*}\NormalTok{ cycle\_length }
\end{Highlighting}
\end{Shaded}

Both treatments A and B incur a cost. To create the vector of state costs for strategy A, \texttt{v\_c\_strA}, we add the cost of treatment A, \texttt{c\_trtA}, to the state costs of S1 and S2. Similarly, when constructing the vector of state costs for strategy B, \texttt{v\_c\_strB}, we add the cost of treatment B, \texttt{c\_trtB}, to the state costs of S1 and S2. Finally, for the vector of state costs for strategy AB, \texttt{v\_c\_strAB}, we add both treatment costs to the state costs of S1 and S2.

\begin{Shaded}
\begin{Highlighting}[]
\CommentTok{\# Vector of state costs for strategy A}
\NormalTok{v\_c\_strA }\OtherTok{\textless{}{-}} \FunctionTok{c}\NormalTok{(}\AttributeTok{H  =}\NormalTok{ c\_H, }
              \AttributeTok{S1 =}\NormalTok{ c\_S1 }\SpecialCharTok{+}\NormalTok{ c\_trtA, }
              \AttributeTok{S2 =}\NormalTok{ c\_S2 }\SpecialCharTok{+}\NormalTok{ c\_trtA, }
              \AttributeTok{D  =}\NormalTok{ c\_D) }\SpecialCharTok{*}\NormalTok{ cycle\_length }
\CommentTok{\# Vector of state costs for strategy B}
\NormalTok{v\_c\_strB }\OtherTok{\textless{}{-}} \FunctionTok{c}\NormalTok{(}\AttributeTok{H  =}\NormalTok{ c\_H, }
              \AttributeTok{S1 =}\NormalTok{ c\_S1 }\SpecialCharTok{+}\NormalTok{ c\_trtB, }
              \AttributeTok{S2 =}\NormalTok{ c\_S2 }\SpecialCharTok{+}\NormalTok{ c\_trtB, }
              \AttributeTok{D  =}\NormalTok{ c\_D) }\SpecialCharTok{*}\NormalTok{ cycle\_length }
\CommentTok{\# Vector of state costs for strategy AB}
\NormalTok{v\_c\_strAB }\OtherTok{\textless{}{-}} \FunctionTok{c}\NormalTok{(}\AttributeTok{H  =}\NormalTok{ c\_H, }
               \AttributeTok{S1 =}\NormalTok{ c\_S1 }\SpecialCharTok{+}\NormalTok{ (c\_trtA }\SpecialCharTok{+}\NormalTok{ c\_trtB), }
               \AttributeTok{S2 =}\NormalTok{ c\_S2 }\SpecialCharTok{+}\NormalTok{ (c\_trtA }\SpecialCharTok{+}\NormalTok{ c\_trtB), }
               \AttributeTok{D  =}\NormalTok{ c\_D) }\SpecialCharTok{*}\NormalTok{ cycle\_length }
\end{Highlighting}
\end{Shaded}

To compute the expected QALYs and costs for the Sick-Sicker model under SoC and strategy A, we apply Equation \eqref{eq:exp-rew-cycle} by multiplying the cohort trace matrix, \texttt{m\_M}, times the corresponding strategy-specific state vectors of rewards. Similarly, to compute the expected rewards for strategies B and AB, we multiply the cohort trace matrix accounting for the effectiveness of treatment B, \texttt{m\_M\_strB}, times their corresponding state vectors of rewards.

\begin{Shaded}
\begin{Highlighting}[]
\CommentTok{\# Vector of QALYs under SoC}
\NormalTok{v\_qaly\_SoC }\OtherTok{\textless{}{-}}\NormalTok{ m\_M }\SpecialCharTok{\%*\%}\NormalTok{ v\_u\_SoC}
\CommentTok{\# Vector of costs under SoC}
\NormalTok{v\_cost\_SoC }\OtherTok{\textless{}{-}}\NormalTok{ m\_M }\SpecialCharTok{\%*\%}\NormalTok{ v\_c\_SoC}
\CommentTok{\# Vector of QALYs for strategy A}
\NormalTok{v\_qaly\_strA }\OtherTok{\textless{}{-}}\NormalTok{ m\_M\_strA }\SpecialCharTok{\%*\%}\NormalTok{ v\_u\_strA}
\CommentTok{\# Vector of costs for strategy A}
\NormalTok{v\_cost\_strA }\OtherTok{\textless{}{-}}\NormalTok{ m\_M\_strA }\SpecialCharTok{\%*\%}\NormalTok{ v\_c\_strA}
\CommentTok{\# Vector of QALYs for strategy B}
\NormalTok{v\_qaly\_strB }\OtherTok{\textless{}{-}}\NormalTok{ m\_M\_strB }\SpecialCharTok{\%*\%}\NormalTok{ v\_u\_strB}
\CommentTok{\# Vector of costs for strategy B}
\NormalTok{v\_cost\_strB }\OtherTok{\textless{}{-}}\NormalTok{ m\_M\_strB }\SpecialCharTok{\%*\%}\NormalTok{ v\_c\_strB}
\CommentTok{\# Vector of QALYs for strategy AB}
\NormalTok{v\_qaly\_strAB }\OtherTok{\textless{}{-}}\NormalTok{ m\_M\_strAB }\SpecialCharTok{\%*\%}\NormalTok{ v\_u\_strAB}
\CommentTok{\# Vector of costs for strategy AB}
\NormalTok{v\_cost\_strAB }\OtherTok{\textless{}{-}}\NormalTok{ m\_M\_strAB }\SpecialCharTok{\%*\%}\NormalTok{ v\_c\_strAB}
\end{Highlighting}
\end{Shaded}

\hypertarget{within-cycle-correction}{%
\subsection{Within-cycle correction}\label{within-cycle-correction}}

A discrete-time cSTM involves an approximation of continuous-time dynamics to discrete points in time. The discretization might introduce biases when estimating outcomes based on state occupancy.\textsuperscript{\protect\hyperlink{ref-VanRosmalen2013}{21}} One approach to reducing these biases is to shorten the cycle length, requiring simulating the model for a larger number of cycles, which can be computationally burdensome. Another approach is to use within-cycle corrections (WCC).\textsuperscript{\protect\hyperlink{ref-Siebert2012c}{2},\protect\hyperlink{ref-Hunink2014}{22}} In this tutorial, we use Simpson's 1/3rd rule by multiplying the rewards (e.g., costs and effectiveness) by \(1/3\) in the first and last cycles, by \(4/3\) for the odd cycles, and by \(2/3\) for the even cycles.\textsuperscript{\protect\hyperlink{ref-Elbasha2016}{23},\protect\hyperlink{ref-Elbasha2016a}{24}} We implement the WCC by generating a column vector \(\mathbf{wcc}\) of size \(n_T+1\) with values corresponding to the first, \(t=0\), and last cycle, \(t= n_T\), equal to \(1/3\), and the entries corresponding to the even and odd cycles with \(2/3\) and \(4/3\), respectively.

\[
  \mathbf{wcc} = \left[\frac{1}{3}, \frac{2}{3}, \frac{4}{3}, \cdots, \frac{1}{3}\right]
\]

Since the WCC vector is the same for costs and QALYs, we only require one vector (\texttt{v\_wcc}). We create \texttt{v\_wcc} by defining two indicator functions that tell us whether the vector entries are even or odd, filled with the corresponding factors given by Simpson's 1/3rd rule. We used the \texttt{function} command that reads a vector \texttt{x} and applies the modulo operation \texttt{\%\%} that returns the remainder of dividing each of the vector entries by 2. If the remainder of the \(i-th\) entry is 0, the entry is even, or it is odd if the remainder is 1 otherwise.

\begin{Shaded}
\begin{Highlighting}[]
\CommentTok{\# First, we define two functions to identify if a number is even or odd}
\NormalTok{is\_even }\OtherTok{\textless{}{-}} \ControlFlowTok{function}\NormalTok{(x) x }\SpecialCharTok{\%\%} \DecValTok{2} \SpecialCharTok{==} \DecValTok{0}
\NormalTok{is\_odd  }\OtherTok{\textless{}{-}} \ControlFlowTok{function}\NormalTok{(x) x }\SpecialCharTok{\%\%} \DecValTok{2} \SpecialCharTok{!=} \DecValTok{0}
\DocumentationTok{\#\# Vector with cycles}
\NormalTok{v\_cycles }\OtherTok{\textless{}{-}} \FunctionTok{seq}\NormalTok{(}\DecValTok{1}\NormalTok{, n\_cycles }\SpecialCharTok{+} \DecValTok{1}\NormalTok{)}
\DocumentationTok{\#\# Generate 2/3 and 4/3 multipliers for even and odd entries, respectively}
\NormalTok{v\_wcc }\OtherTok{\textless{}{-}} \FunctionTok{is\_even}\NormalTok{(v\_cycles)}\SpecialCharTok{*}\NormalTok{(}\DecValTok{2}\SpecialCharTok{/}\DecValTok{3}\NormalTok{) }\SpecialCharTok{+} \FunctionTok{is\_odd}\NormalTok{(v\_cycles)}\SpecialCharTok{*}\NormalTok{(}\DecValTok{4}\SpecialCharTok{/}\DecValTok{3}\NormalTok{)}
\DocumentationTok{\#\# Substitute 1/3 in first and last entries}
\NormalTok{v\_wcc[}\DecValTok{1}\NormalTok{] }\OtherTok{\textless{}{-}}\NormalTok{ v\_wcc[n\_cycles }\SpecialCharTok{+} \DecValTok{1}\NormalTok{] }\OtherTok{\textless{}{-}} \DecValTok{1}\SpecialCharTok{/}\DecValTok{3}
\end{Highlighting}
\end{Shaded}

\hypertarget{discounting-future-rewards}{%
\subsection{Discounting future rewards}\label{discounting-future-rewards}}

We often discount future costs and benefits by a specific rate to calculate the net present value of these rewards. We then use this rate to generate a column vector with cycle-specific discount weights \(\mathbf{d}\) of size \(n_T+1\) where its \(t\)-th entry represents the discounting for cycle \(t\)

\[
  \mathbf{d} = \left[1, \frac{1}{(1+d)^{1}}, \frac{1}{(1+d)^{2}}, \cdots, \frac{1}{(1+d)^{n_T}}\right],
\]
where \(d\) is the cycle-length discount rate. At the end of the simulation, we multiply the vector of expected rewards, \(\mathbf{y}\), by a discounting column vector. The total expected discounted outcome summed over the \(n_T\) cycles, \(y\), is obtained by the inner product between \(\mathbf{y}\) transposed, \(\mathbf{y'}\), and \(\mathbf{d}\),
\begin{equation}
 y = \mathbf{y'} \mathbf{d}.
 \label{eq:tot-exp-disc-rewd}
\end{equation}

The discount vectors for costs and QALYs for the Sick-Sicker model with annual cycles, \texttt{v\_dwc} and \texttt{v\_dwe}, respectively, scaled by the cycle length, are

\begin{Shaded}
\begin{Highlighting}[]
\CommentTok{\# Discount weight for effects}
\NormalTok{v\_dwe }\OtherTok{\textless{}{-}} \DecValTok{1} \SpecialCharTok{/}\NormalTok{ ((}\DecValTok{1} \SpecialCharTok{+}\NormalTok{ (d\_e }\SpecialCharTok{*}\NormalTok{ cycle\_length)) }\SpecialCharTok{\^{}}\NormalTok{ (}\DecValTok{0}\SpecialCharTok{:}\NormalTok{n\_cycles))  }
\CommentTok{\# Discount weight for costs }
\NormalTok{v\_dwc }\OtherTok{\textless{}{-}} \DecValTok{1} \SpecialCharTok{/}\NormalTok{ ((}\DecValTok{1} \SpecialCharTok{+}\NormalTok{ (d\_c }\SpecialCharTok{*}\NormalTok{ cycle\_length)) }\SpecialCharTok{\^{}}\NormalTok{ (}\DecValTok{0}\SpecialCharTok{:}\NormalTok{n\_cycles))    }
\end{Highlighting}
\end{Shaded}

The functions and code for creating the WCC and discounting vectors above are single lines of code that affect the entire vectors of rewards used to compute the health and economic outputs of the model. To compute the total expected discounted QALYs and costs under all four strategies accounting for both discounting and WCC, we incorporate \(\mathbf{wcc}\) in equation \eqref{eq:tot-exp-disc-rewd} using an element-wise multiplication with \(\mathbf{d}\), indicated by the \(\odot\) sign. The element-wise multiplication computes a new vector with elements that are the products of the corresponding elements of \(\mathbf{wcc}\) and \(\mathbf{d}\).
\begin{equation}
 y = \mathbf{y}^{'} \left(\mathbf{d} \odot \mathbf{wcc}\right).
 \label{eq:tot-exp-disc-rewd-wcc}
\end{equation}

To compute the total expected discounted and WCC-corrected QALYs under all four strategies in R, we apply Equation \eqref{eq:tot-exp-disc-rewd-wcc} to the reward vectors of each strategy.

\begin{Shaded}
\begin{Highlighting}[]
\DocumentationTok{\#\# Expected discounted QALYs under SoC}
\NormalTok{n\_tot\_qaly\_SoC }\OtherTok{\textless{}{-}} \FunctionTok{t}\NormalTok{(v\_qaly\_SoC) }\SpecialCharTok{\%*\%}\NormalTok{ (v\_dwe }\SpecialCharTok{*}\NormalTok{ v\_wcc)}
\DocumentationTok{\#\# Expected discounted costs under SoC}
\NormalTok{n\_tot\_cost\_SoC }\OtherTok{\textless{}{-}} \FunctionTok{t}\NormalTok{(v\_cost\_SoC) }\SpecialCharTok{\%*\%}\NormalTok{ (v\_dwc }\SpecialCharTok{*}\NormalTok{ v\_wcc)}
\DocumentationTok{\#\# Expected discounted QALYs for strategy A}
\NormalTok{n\_tot\_qaly\_strA }\OtherTok{\textless{}{-}} \FunctionTok{t}\NormalTok{(v\_qaly\_strA) }\SpecialCharTok{\%*\%}\NormalTok{ (v\_dwe }\SpecialCharTok{*}\NormalTok{ v\_wcc)}
\DocumentationTok{\#\# Expected discounted costs for strategy A}
\NormalTok{n\_tot\_cost\_strA }\OtherTok{\textless{}{-}} \FunctionTok{t}\NormalTok{(v\_cost\_strA) }\SpecialCharTok{\%*\%}\NormalTok{ (v\_dwc }\SpecialCharTok{*}\NormalTok{ v\_wcc)}
\DocumentationTok{\#\# Expected discounted QALYs for strategy B}
\NormalTok{n\_tot\_qaly\_strB }\OtherTok{\textless{}{-}} \FunctionTok{t}\NormalTok{(v\_qaly\_strB) }\SpecialCharTok{\%*\%}\NormalTok{ (v\_dwe }\SpecialCharTok{*}\NormalTok{ v\_wcc)}
\DocumentationTok{\#\# Expected discounted costs for strategy B}
\NormalTok{n\_tot\_cost\_strB }\OtherTok{\textless{}{-}} \FunctionTok{t}\NormalTok{(v\_cost\_strB) }\SpecialCharTok{\%*\%}\NormalTok{ (v\_dwc }\SpecialCharTok{*}\NormalTok{ v\_wcc)}
\DocumentationTok{\#\# Expected discounted QALYs for strategy AB}
\NormalTok{n\_tot\_qaly\_strAB }\OtherTok{\textless{}{-}} \FunctionTok{t}\NormalTok{(v\_qaly\_strAB) }\SpecialCharTok{\%*\%}\NormalTok{ (v\_dwe }\SpecialCharTok{*}\NormalTok{ v\_wcc)}
\DocumentationTok{\#\# Expected discounted costs for strategy AB}
\NormalTok{n\_tot\_cost\_strAB }\OtherTok{\textless{}{-}} \FunctionTok{t}\NormalTok{(v\_cost\_strAB) }\SpecialCharTok{\%*\%}\NormalTok{ (v\_dwc }\SpecialCharTok{*}\NormalTok{ v\_wcc)}
\end{Highlighting}
\end{Shaded}

\begin{table}[!h]

\caption{\label{tab:Expected-outcomes-table}Total expected discounted QALYs and costs per average individual in the cohort of the Sick-Sicker model by strategy accounting for within-cycle correction.}
\centering
\begin{tabular}[t]{llc}
\toprule{}
  & Costs & QALYs\\
\midrule{}
Standard of care & \$151,580 & 20.711\\
Strategy A & \$284,805 & 21.499\\
Strategy B & \$259,100 & 22.184\\
Strategy AB & \$378,875 & 23.137\\
\bottomrule{}
\end{tabular}
\end{table}

The total expected discounted QALYs and costs for the Sick-Sicker model under the four strategies accounting for within-cycle correction are shown in Table \ref{tab:Expected-outcomes-table}.

\hypertarget{incremental-cost-effectiveness-ratios-icers}{%
\section{Incremental cost-effectiveness ratios (ICERs)}\label{incremental-cost-effectiveness-ratios-icers}}

We combine the total expected discounted costs and QALYs for all four strategies into outcome-specific vectors, \texttt{v\_cost\_str} for costs and \texttt{v\_qaly\_str} for QALYs. So far, we have used \emph{base} R to create and simulate cSTMs. For the CEA, we use the various functions from R package \texttt{dampack} (\url{https://cran.r-project.org/web/packages/dampack/})\textsuperscript{\protect\hyperlink{ref-Alarid-Escudero2021}{25}} that are also included as supplementary material to calculate the incremental costs and effectiveness and the incremental cost-effectiveness ratio (ICER) between non-dominated strategies and create the data frame \texttt{df\_cea} with this information. These outcomes are required inputs to conduct a CEA. We included function from `dampack` for the probabilistic sensitivity analysis (PSA) below.

\begin{Shaded}
\begin{Highlighting}[]
\DocumentationTok{\#\#\# Vector of costs}
\NormalTok{v\_cost\_str }\OtherTok{\textless{}{-}} \FunctionTok{c}\NormalTok{(n\_tot\_cost\_SoC, n\_tot\_cost\_strA, n\_tot\_cost\_strB, n\_tot\_cost\_strAB)}
\DocumentationTok{\#\#\# Vector of effectiveness}
\NormalTok{v\_qaly\_str }\OtherTok{\textless{}{-}} \FunctionTok{c}\NormalTok{(n\_tot\_qaly\_SoC, n\_tot\_qaly\_strA, n\_tot\_qaly\_strB, n\_tot\_qaly\_strAB)}

\DocumentationTok{\#\#\# Calculate incremental cost{-}effectiveness ratios (ICERs)}
\NormalTok{df\_cea }\OtherTok{\textless{}{-}}\NormalTok{ dampack}\SpecialCharTok{::}\FunctionTok{calculate\_icers}\NormalTok{(}\AttributeTok{cost =}\NormalTok{ v\_cost\_str, }
                                   \AttributeTok{effect =}\NormalTok{ v\_qaly\_str,}
                                   \AttributeTok{strategies =}\NormalTok{ v\_names\_str)}
\end{Highlighting}
\end{Shaded}

\begin{figure}[H]

{\centering \includegraphics{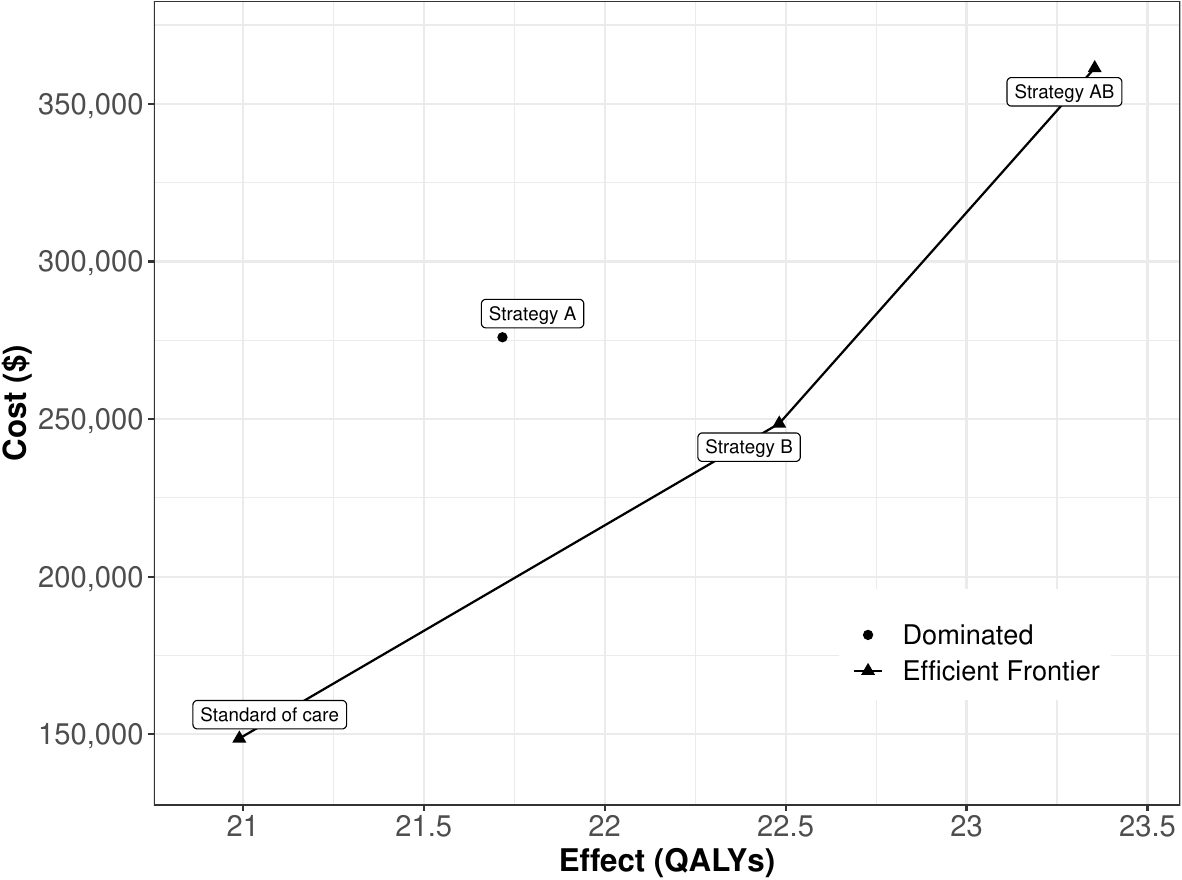} 

}

\caption{Cost-effectiveness efficient frontier of the cost-effectiveness analysis based on the time-independent Sick-Sicker model.}\label{fig:Sick-Sicker-CEA}
\end{figure}

SoC is the least costly and effective strategy, followed by Strategy B producing an expected incremental benefit of 1.473 QALYs per individual for an additional expected cost of \$107,521 with an ICER of \$72,988/QALY followed by Strategy AB with an ICER \$125,764/QALY. Strategy A is a dominated strategy (Table \ref{tab:table-cea}). Strategies SoC, B and AB form the cost-effectiveness efficient frontier (Figure \ref{fig:Sick-Sicker-CEA}).

\begin{table}[!h]

\caption{\label{tab:table-cea}Cost-effectiveness analysis results for the Sick-Sicker model. ND: Non-dominated strategy; D: Dominated strategy.}
\centering
\begin{tabular}[t]{rcccccc}
\toprule{}
Strategy & Costs (\$) & QALYs & Incremental Costs (\$) & Incremental QALYs & ICER (\$/QALY) & Status\\
\midrule{}
Standard of care & 151,580 & 20.711 & NA & NA & NA & ND\\
Strategy B & 259,100 & 22.184 & 107,521 & 1.473 & 72,988 & ND\\
Strategy AB & 378,875 & 23.137 & 119,775 & 0.952 & 125,764 & ND\\
Strategy A & 284,805 & 21.499 & NA & NA & NA & D\\
\bottomrule{}
\end{tabular}
\end{table}

\hypertarget{probabilistic-sensitivity-analysis}{%
\section{Probabilistic sensitivity analysis}\label{probabilistic-sensitivity-analysis}}

To quantify the effect of model parameter uncertainty on cost-effectiveness outcomes, we conducted a PSA by randomly drawing \(K\) parameter sets (\texttt{n\_sim}) from distributions that reflect the current uncertainty in model parameter estimates.\textsuperscript{\protect\hyperlink{ref-Briggs2012}{26}} The distribution for all the parameters and their values are described in Table \ref{tab:param-table} and in more detail in the Supplementary Material. We compute model outcomes for each sampled set of parameter values (e.g., total discounted cost and QALYs) for each strategy. In a previously published manuscript, we describe the implementation of these steps in R.\textsuperscript{\protect\hyperlink{ref-Alarid-Escudero2019e}{20}} Briefly, to conduct the PSA, we create three R functions:

\begin{enumerate}
\def\labelenumi{\arabic{enumi}.}
\item
  \texttt{generate\_psa\_params(n\_sim,\ seed)}: a function that generates a sample of size \texttt{n\_sim} for the model parameters, \texttt{df\_psa\_input}, from their distributions defined in Table \ref{tab:param-table}. The function input \texttt{seed} sets the seed of the pseudo-random number generator used in sampling parameter values, which ensures reproducibility of the PSA results.
\item
  \texttt{decision\_model(l\_params\_all,\ verbose\ =\ FALSE)}: a function that wraps the R code of the time-independent cSTM described in section \ref{time-independent-cstm-dynamics}. This function requires inputting a list of all model parameter values, \texttt{l\_params\_all} and whether the user wants print messages on whether transition probability matrices are valid via the \texttt{verbose} parameter.
\item
  \texttt{calculate\_ce\_out(l\_params\_all,\ n\_wtp\ =\ 100000)}: a function that calculates total discounted costs and QALYs based on the \texttt{decision\_model} function output. This function also computes the net monetary benefit (NMB) for a given willingness-to-pay threshold, specified by the argument \texttt{n\_wtp}.
  These functions are provided in the accompanying GitHub repository of this manuscript.
\end{enumerate}

\begin{figure}[H]

{\centering \includegraphics{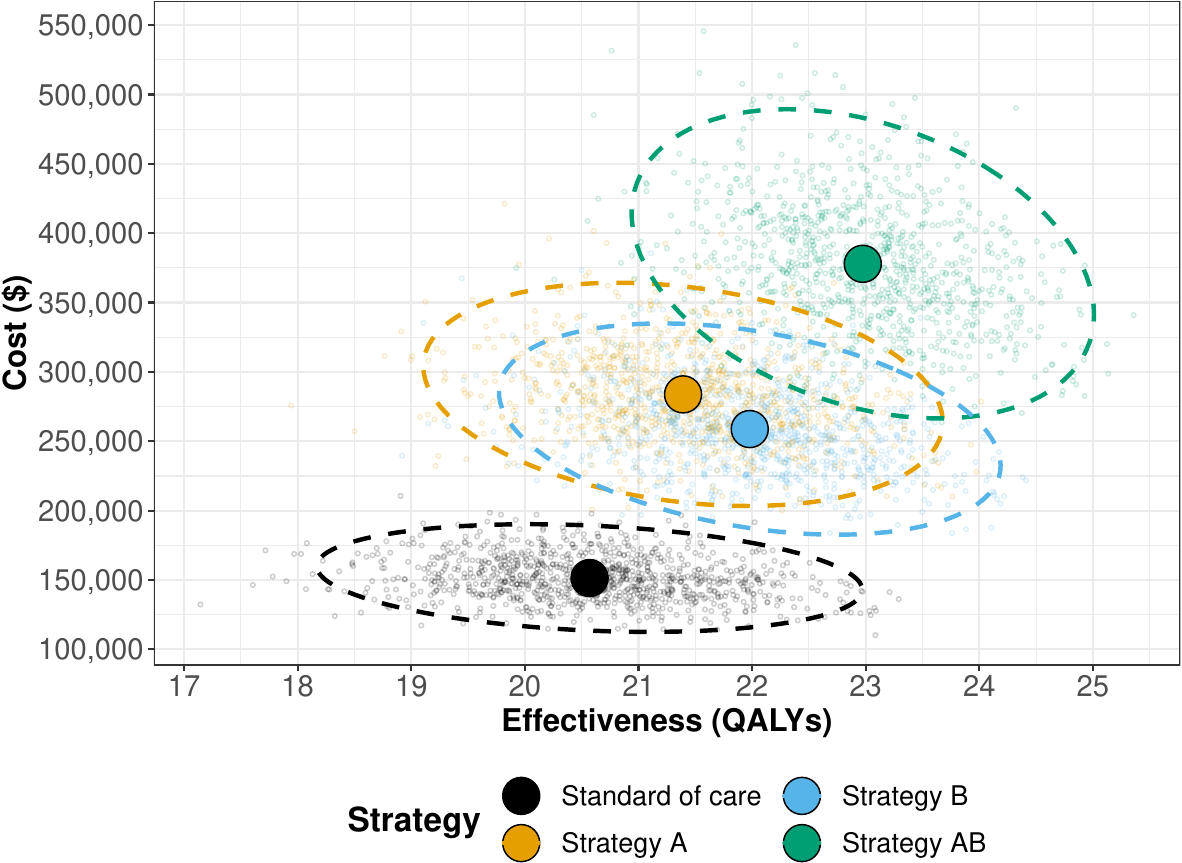} 

}

\caption{Cost-effectiveness scatter plot.}\label{fig:CE-scatter}
\end{figure}

To conduct the PSA of the CEA using the time-independent Sick-Sicker cSTM, we sampled 1,000 parameter sets from their distributions. We assumed commonly used distributions to describe their uncertainty for each type of parameter. For example, gamma for transition rates, lognormal for hazard ratios, and beta for utility weights.\textsuperscript{\protect\hyperlink{ref-Hunink2014}{22},\protect\hyperlink{ref-Parmigiani1997}{27}--\protect\hyperlink{ref-Briggs2002}{29}} For each sampled parameter set, we simulated the cost and effectiveness of each strategy. Results from a PSA can be represented in various ways. For example, the joint distribution, 95\% confidence ellipse, and the expected values of the total discounted costs and QALYs for each strategy can be plotted in a cost-effectiveness scatter plot (Figure \ref{fig:CE-scatter}),\textsuperscript{\protect\hyperlink{ref-Briggs2002}{29}} where each of the 1,000 simulations are plotted as a point in the graph. The CE scatter plot for CEA using the time-independent model shows that strategy AB has the highest expected costs and QALYs. Standard of care has the lowest expected cost and QALYs. Strategy B is more effective and least costly than Strategy A. Strategy A is a strongly dominated strategy.

\begin{figure}[H]

{\centering \includegraphics{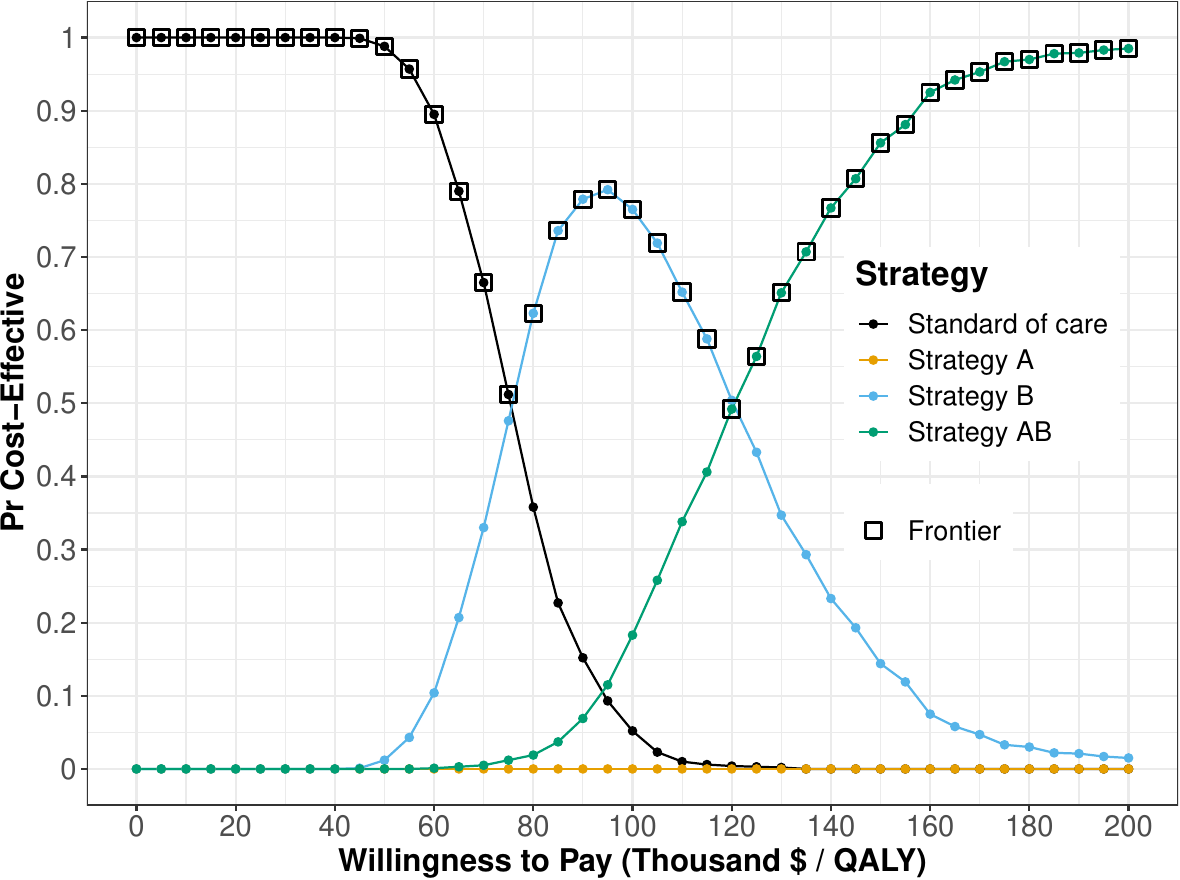} 

}

\caption{Cost-effectiveness acceptability curves (CEACs) and frontier (CEAF).}\label{fig:CEAC}
\end{figure}

Figure \ref{fig:CEAC} presents the cost-effectiveness acceptability curves (CEACs), which show the probability that each strategy is cost-effective, and the cost-effectiveness frontier (CEAF), which shows the strategy with the highest expected net monetary benefit (NMB), over a range of willingness-to-pay (WTP) thresholds. Each strategy's NMB is computed using \(\text{NMB} = \text{QALY} \times \text{WTP} - \text{Cost}\)\textsuperscript{\protect\hyperlink{ref-Stinnett1998b}{30}} for each PSA sample. At WTP thresholds less than \$80,000 per QALY gained, strategy SoC has both the highest probability of being cost-effective and the highest expected NMB. This switches to strategy B for WTP thresholds between \$80,000 and \$120,000 per QALY gained and to strategy AB for WTP thresholds greater than or equal to \$120,000 per QALY gained.

\begin{figure}[H]

{\centering \includegraphics{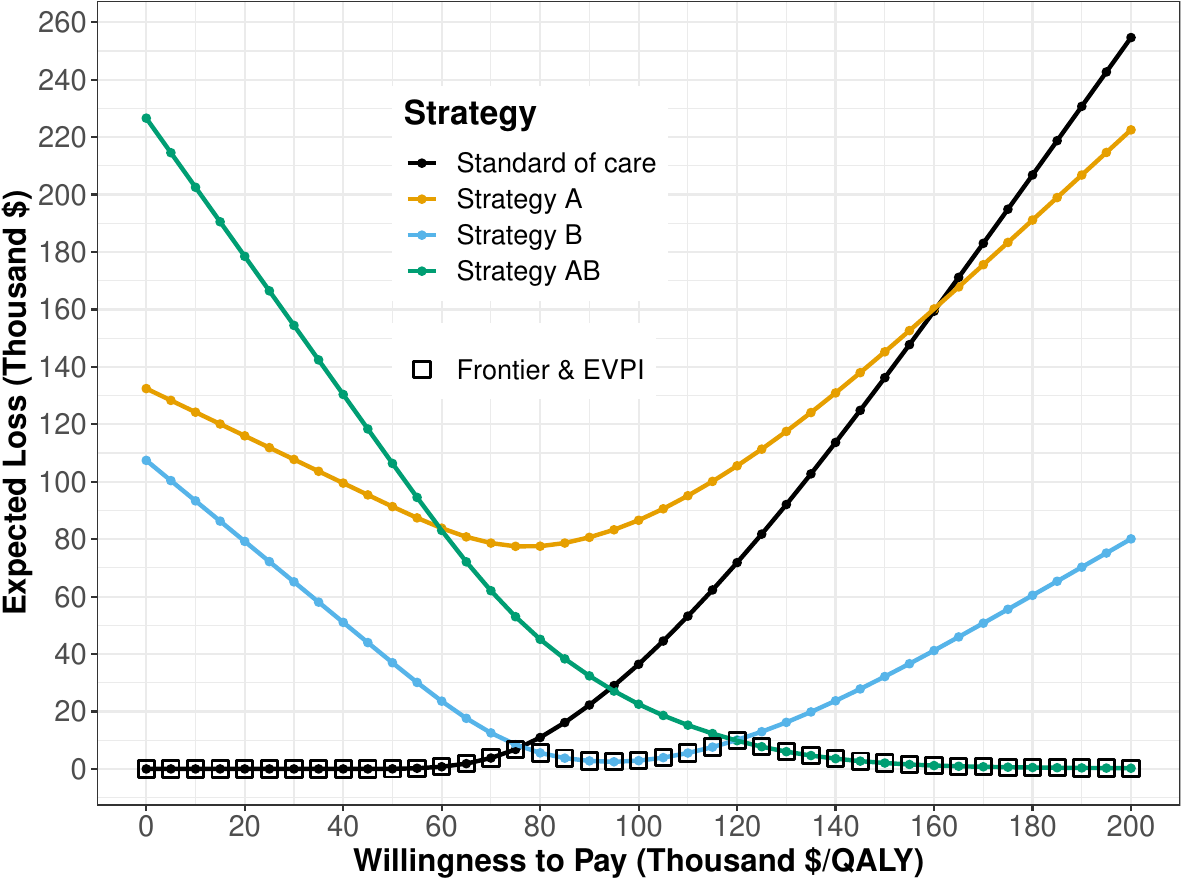} 

}

\caption{Expected loss curves (ELCs) and expected value of perfect information (EVPI).}\label{fig:ELC}
\end{figure}

The CEAC and CEAF do not show the magnitude of the expected net benefit lost (i.e., expected loss) when the chosen strategy is not the cost-effective strategy in all the samples of the PSA. To complement these results, we quantify expected loss from each strategy over a range of WTP thresholds with the expected loss curves (ELCs). These curves quantify the expected loss from each strategy over a range of WTP thresholds (Figure \ref{fig:ELC}). The expected loss considers both the probability of making the wrong decision and the magnitude of the loss due to this decision, representing the foregone benefits of choosing a suboptimal strategy. The expected loss of the optimal strategy represents the lowest envelope of the ELCs because, given current information, the loss cannot be minimized further. The lower bound of the ELCs represents the expected value of perfect information (EVPI), which quantifies the value of eliminating parameter uncertainty. We refer the reader to previously published literature for a more detailed description of CEAC, CEAF, and ELC interpretations and the R code to generate them.\textsuperscript{\protect\hyperlink{ref-Alarid-Escudero2019}{31}}

\hypertarget{discussion}{%
\section{Discussion}\label{discussion}}

In this introductory tutorial, we provided a step-by-step mathematical conceptualization of time-independent cSTMs and a walk-through of their implementation in R using a hypothetical disease example with accompanying code throughout the tutorial. While some of the presented implementation details are specific to the R programming language, much of the code structure shown in this tutorial would be similar in other programming languages. Thus, readers may use this tutorial as a template for coding cSTMs more generally in different programming languages.

The parameterization of our example model assumes all parameters are known, or at least, the characterization of their uncertainty is known (i.e., we know their distributions). However, to construct a real-world cSTM, modelers must conduct a thorough synthesis of current evidence to determine these appropriate structures and inform all parameters based on the current evidence. For example, literature must be carefully considered when determining whether transitions between non-death health states are estimated conditional on being alive or are estimated as competing risks along with mortality risks.\textsuperscript{\protect\hyperlink{ref-Briggs2012}{26}} Similarly, our PSA analysis simplifies reality where all model parameters are assumed to be independent of each other. However, parameters could be correlated or have a rank order, and appropriate statistical methods that simulate these correlations or rank order might be needed.\textsuperscript{\protect\hyperlink{ref-Goldhaber-Fiebert2015}{32}} We encourage modelers to use appropriate statistical methods to accurately synthesize and quantify model parameters' uncertainty. For example, for the PSA of our case study, we used distributions based on the type of parameters following standard recommendations. For a more detailed description of how to choose distributions, we refer the reader to other literature.\textsuperscript{\protect\hyperlink{ref-Briggs2002}{29},\protect\hyperlink{ref-Briggs2003}{33}} In addition, modelers should appropriately specify all model parameters for the cycle length of the model.\textsuperscript{\protect\hyperlink{ref-Hunink2014}{22}}

In general, cSTMs are recommended when the number of states is considered ``not too large.''\textsuperscript{\protect\hyperlink{ref-Siebert2012c}{2}} This recommendation arises because it becomes more challenging to keep track of their construction as the number of states increases. It is possible to build reasonably complex cSTMs in R as long as the computer's RAM can store the size of the transition probability matrix and outputs of interest. For time-independent cSTMs, in general, this should not be a problem with the capacity of current RAM in personal computers. An alternative to reduce the explosion of disease states is individual-based state-transition model (iSTM), which is a type of STM where simulated individuals transition between health states over time.\textsuperscript{\protect\hyperlink{ref-Siebert2012c}{2}} We have previously published a tutorial on the implementation of iSTM in R.\textsuperscript{\protect\hyperlink{ref-Krijkamp2018}{19}}

With increasing model complexity and interdependency of functions to conduct various analyses like PSA, it is essential to ensure all code and functions work as expected and all elements of the cSTM are valid. We can achieve this by creating functions that help with model debugging, validation, and thorough unit testing. In the accompanying GitHub repository, we provide functions to check that transition probability matrices and their elements are valid. These functions are an example of a broader standard practice in software development called unit testing that requires building functions to test and check that the model and model-based analysis perform as intended.\textsuperscript{\protect\hyperlink{ref-Wickham2021}{34}} However, unit testing is beyond the scope of this tutorial. We refer the reader to a previously published manuscript that describes unit testing in more detail and provides accompanying code.\textsuperscript{\protect\hyperlink{ref-Alarid-Escudero2019e}{20}}

In this tutorial, we implemented a cSTM using a (discrete-time) transition matrix. However, cSTM can also be implemented via (discrete-time) difference equations or (continuous-time) differential equations in R.\textsuperscript{\protect\hyperlink{ref-Grimmett2014}{35},\protect\hyperlink{ref-Axler2005}{36}} We refer readers interested in learning more on continuous-time cSTMs to previously published manuscripts\textsuperscript{\protect\hyperlink{ref-VanRosmalen2013}{21},\protect\hyperlink{ref-Cao2016}{37}--\protect\hyperlink{ref-Soares2012}{39}} and a tutorial using R.\textsuperscript{\protect\hyperlink{ref-Frederix2013a}{40}} Finally, the variable names used in this paper reflect our coding style. While we provide standardized variable names, adopting these conventions is ultimately a personal preference.

In summary, this tutorial provides a conceptualization of time-independent cSTMs and a step-by-step guide to implement them in R. We aim to add to the current body of literature and material on building this type of decision model so that health decision scientists and health economists can develop cSTMs in a more flexible, efficient, open-source manner and to encourage increased transparency and reproducibility. In the advanced cSTM tutorial, we explore generalizing this framework to time-dependent cSTM, generating epidemiological outcomes, and incorporating transition rewards.

\section*{Acknowledgements}\label{acknowledgements}
\addcontentsline{toc}{section}{Acknowledgements}

Dr.~Alarid-Escudero was supported by grants U01-CA199335 and U01-CA253913 from the National Cancer Institute (NCI) as part of the Cancer Intervention and Surveillance Modeling Network (CISNET), and a grant by the Gordon and Betty Moore Foundation. Miss Krijkamp was supported by the Society for Medical Decision Making (SMDM) fellowship through a grant by the Gordon and Betty Moore Foundation (GBMF7853). Dr.~Enns was supported by a grant from the National Institute of Allergy and Infectious Diseases of the National Institutes of Health under award no. K25AI118476. Dr.~Hunink received research funding from the American Diabetes Association, the Netherlands Organization for Health Research and Development, the German Innovation Fund, Netherlands Educational Grant (``Studie Voorschot Middelen''), and the Gordon and Betty Moore Foundation. Dr.~Jalal was supported by a grant from the National Institute on Drug Abuse of the National Institute of Health under award no. K01DA048985. The content is solely the responsibility of the authors and does not necessarily represent the official views of the National Institutes of Health. The funding agencies had no role in the design of the study, interpretation of results, or writing of the manuscript. The funding agreement ensured the authors' independence in designing the study, interpreting the data, writing, and publishing the report. We also want to thank the anonymous reviewers of \emph{Medical Decision Making} for their valuable suggestions and the students who took our classes where we refined these materials.

\hypertarget{references}{%
\section*{References}\label{references}}
\addcontentsline{toc}{section}{References}

\hypertarget{refs}{}
\begin{CSLReferences}{0}{0}
\leavevmode\hypertarget{ref-Kuntz2017}{}%
\CSLLeftMargin{1. }
\CSLRightInline{Kuntz KM, Russell LB, Owens DK, Sanders GD, Trikalinos TA, Salomon JA. {Decision Models in Cost-Effectiveness Analysis}. In: Neumann PJ, Sanders GD, Russell LB, Siegel JE, Ganiats TG, editors. Cost-effectiveness in health and medicine. Second. New York, NY: Oxford University Press; 2017. p. 105--36. }

\leavevmode\hypertarget{ref-Siebert2012c}{}%
\CSLLeftMargin{2. }
\CSLRightInline{Siebert U, Alagoz O, Bayoumi AM, Jahn B, Owens DK, Cohen DJ, et al. {State-Transition Modeling: A Report of the ISPOR-SMDM Modeling Good Research Practices Task Force-3}. Medical Decision Making {[}Internet{]}. 2012;32(5):690--700. Available from: \url{http://mdm.sagepub.com/cgi/doi/10.1177/0272989X12455463}}

\leavevmode\hypertarget{ref-Suijkerbuijk2018}{}%
\CSLLeftMargin{3. }
\CSLRightInline{Suijkerbuijk AWM, Van Hoek AJ, Koopsen J, De Man RA, Mangen MJJ, De Melker HE, et al. {Cost-effectiveness of screening for chronic hepatitis B and C among migrant populations in a low endemic country}. PLoS ONE. 2018;13(11):1--6. }

\leavevmode\hypertarget{ref-Sathianathen2018a}{}%
\CSLLeftMargin{4. }
\CSLRightInline{Sathianathen NJ, Konety BR, Alarid-Escudero F, Lawrentschuk N, Bolton DM, Kuntz KM. {Cost-effectiveness Analysis of Active Surveillance Strategies for Men with Low-risk Prostate Cancer}. European Urology {[}Internet{]}. 2019;75(6):910--7. Available from: \url{https://linkinghub.elsevier.com/retrieve/pii/S0302283818308534}}

\leavevmode\hypertarget{ref-Lu2018b}{}%
\CSLLeftMargin{5. }
\CSLRightInline{Lu S, Yu Y, Fu S, Ren H. {Cost-effectiveness of ALK testing and first-line crizotinib therapy for non-small-cell lung cancer in China}. PLoS ONE. 2018;13(10):1--2. }

\leavevmode\hypertarget{ref-Djatche2018}{}%
\CSLLeftMargin{6. }
\CSLRightInline{Djatche LM, Varga S, Lieberthal RD. {Cost-Effectiveness of Aspirin Adherence for Secondary Prevention of Cardiovascular Events}. PharmacoEconomics - Open {[}Internet{]}. 2018;2(4):371--80. Available from: \url{https://doi.org/10.1007/s41669-018-0075-2}}

\leavevmode\hypertarget{ref-Smith-Spangler2010}{}%
\CSLLeftMargin{7. }
\CSLRightInline{Smith-Spangler CM, Juusola JL, Enns EA, Owens DK, Garber AM. {Population Strategies to Decrease Sodium Intake and the Burden of Cardiovascular Disease: A Cost-Effectiveness Analysis}. Annals of Internal Medicine {[}Internet{]}. 2010;152(8):481--7. Available from: \url{http://annals.org/article.aspx?articleid=745729}}

\leavevmode\hypertarget{ref-Pershing2014}{}%
\CSLLeftMargin{8. }
\CSLRightInline{Pershing S, Enns EA, Matesic B, Owens DK, Goldhaber-Fiebert JD. {Cost-Effectiveness of Treatment of Diabetic Macular Edema}. Annals of Internal Medicine. 2014;160(1):18--29. }

\leavevmode\hypertarget{ref-Jalal2017b}{}%
\CSLLeftMargin{9. }
\CSLRightInline{Jalal H, Pechlivanoglou P, Krijkamp E, Alarid-Escudero F, Enns EA, Hunink MGM. {An Overview of R in Health Decision Sciences}. Medical Decision Making {[}Internet{]}. 2017;37(7):735--46. Available from: \url{http://journals.sagepub.com/doi/10.1177/0272989X16686559}}

\leavevmode\hypertarget{ref-Spedicato2017}{}%
\CSLLeftMargin{10. }
\CSLRightInline{Spedicato GA. {Discrete Time Markov Chains with R}. The R Journal {[}Internet{]}. 2017;9(2):84--104. Available from: \url{https://journal.r-project.org/archive/2017/RJ-2017-036/RJ-2017-036.pdf}}

\leavevmode\hypertarget{ref-Filipovic-Pierucci2017}{}%
\CSLLeftMargin{11. }
\CSLRightInline{Filipović-Pierucci A, Zarca K, Durand-Zaleski I. {Markov Models for Health Economic Evaluation: The R Package heemod}. arXiv:170203252v1 {[}Internet{]}. 2017;April:30. Available from: \url{http://arxiv.org/abs/1702.03252}}

\leavevmode\hypertarget{ref-Miller1994}{}%
\CSLLeftMargin{12. }
\CSLRightInline{Miller DK, Homan SM. {Determining Transition Probabilities: Confusion and Suggestions}. Medical Decision Making {[}Internet{]}. 1994 Feb;14(1):52--8. Available from: \url{http://mdm.sagepub.com/cgi/doi/10.1177/0272989X9401400107}}

\leavevmode\hypertarget{ref-Kuntz2001}{}%
\CSLLeftMargin{13. }
\CSLRightInline{Kuntz KM, Weinstein MC. {Modelling in economic evaluation}. In: Drummond MF, McGuire A, editors. Economic evaluation in health care: Merging theory with practice. 2nd ed. New York, NY: Oxford University Press; 2001. p. 141--71. }

\leavevmode\hypertarget{ref-Sonnenberg1993}{}%
\CSLLeftMargin{14. }
\CSLRightInline{Sonnenberg FA, Beck JR. {Markov models in medical decision making: A practical guide}. Medical Decision Making {[}Internet{]}. 1993;13(4):322--38. Available from: \url{http://mdm.sagepub.com/cgi/doi/10.1177/0272989X9301300409}}

\leavevmode\hypertarget{ref-Beck1983}{}%
\CSLLeftMargin{15. }
\CSLRightInline{Beck JR, Pauker SG. {The Markov process in medical prognosis}. Medical Decision Making {[}Internet{]}. 1983;3(4):419--58. Available from: \url{http://mdm.sagepub.com/cgi/doi/10.1177/0272989X8300300403}}

\leavevmode\hypertarget{ref-Alarid-Escudero2021b}{}%
\CSLLeftMargin{16. }
\CSLRightInline{Alarid-Escudero F, Krijkamp E, Enns EA, Yang A, Hunink MGGM, Pechlivanoglou P, et al. {A Tutorial on Time-Dependent Cohort State-Transition Models in R using a Cost-Effectiveness Analysis Example}. arXiv:210813552v1 {[}Internet{]}. 2021;1--41. Available from: \url{https://arxiv.org/abs/2108.13552}}

\leavevmode\hypertarget{ref-Iskandar2018a}{}%
\CSLLeftMargin{17. }
\CSLRightInline{Iskandar R. {A theoretical foundation of state-transition cohort models in health decision analysis}. PLOS ONE {[}Internet{]}. 2018;13(12):e0205543. Available from: \url{https://www.biorxiv.org/content/early/2018/09/28/430173}}

\leavevmode\hypertarget{ref-Enns2015e}{}%
\CSLLeftMargin{18. }
\CSLRightInline{Enns EA, Cipriano LE, Simons CT, Kong CY. {Identifying Best-Fitting Inputs in Health-Economic Model Calibration: A Pareto Frontier Approach}. Medical Decision Making {[}Internet{]}. 2015;35(2):170--82. Available from: \url{http://www.ncbi.nlm.nih.gov/pubmed/24799456}}

\leavevmode\hypertarget{ref-Krijkamp2018}{}%
\CSLLeftMargin{19. }
\CSLRightInline{Krijkamp EM, Alarid-Escudero F, Enns EA, Jalal HJ, Hunink MGM, Pechlivanoglou P. {Microsimulation Modeling for Health Decision Sciences Using R: A Tutorial}. Medical Decision Making {[}Internet{]}. 2018 Apr;38(3):400--22. Available from: \url{http://journals.sagepub.com/doi/10.1177/0272989X18754513}}

\leavevmode\hypertarget{ref-Alarid-Escudero2019e}{}%
\CSLLeftMargin{20. }
\CSLRightInline{Alarid-Escudero F, Krijkamp E, Pechlivanoglou P, Jalal H, Kao S-YZ, Yang A, et al. {A Need for Change! A Coding Framework for Improving Transparency in Decision Modeling}. PharmacoEconomics {[}Internet{]}. 2019;37(11):1329--39. Available from: \url{https://doi.org/10.1007/s40273-019-00837-x}}

\leavevmode\hypertarget{ref-VanRosmalen2013}{}%
\CSLLeftMargin{21. }
\CSLRightInline{Rosmalen J van, Toy M, O'Mahony JF. {A Mathematical Approach for Evaluating Markov Models in Continuous Time without Discrete-Event Simulation}. Medical Decision Making {[}Internet{]}. 2013 May;33(6):767--79. Available from: \url{http://mdm.sagepub.com/cgi/doi/10.1177/0272989X13487947}}

\leavevmode\hypertarget{ref-Hunink2014}{}%
\CSLLeftMargin{22. }
\CSLRightInline{Hunink MGGM, Weinstein MC, Wittenberg E, Drummond MF, Pliskin JS, Wong JB, et al. {Decision Making in Health and Medicine} {[}Internet{]}. 2nd ed. Cambridge: Cambridge University Press; 2014. Available from: \url{http://ebooks.cambridge.org/ref/id/CBO9781139506779}}

\leavevmode\hypertarget{ref-Elbasha2016}{}%
\CSLLeftMargin{23. }
\CSLRightInline{Elbasha EH, Chhatwal J. {Theoretical foundations and practical applications of within-cycle correction methods}. Medical Decision Making. 2016;36(1):115--31. }

\leavevmode\hypertarget{ref-Elbasha2016a}{}%
\CSLLeftMargin{24. }
\CSLRightInline{Elbasha EH, Chhatwal J. {Myths and misconceptions of within-cycle correction: a guide for modelers and decision makers}. PharmacoEconomics. 2016;34(1):13--22. }

\leavevmode\hypertarget{ref-Alarid-Escudero2021}{}%
\CSLLeftMargin{25. }
\CSLRightInline{Alarid-Escudero F, Knowlton G, Easterly CA, Enns EA. Decision analytic modeling package (dampack) {[}Internet{]}. 2021. Available from: \url{https://cran.r-project.org/web/packages/dampack/\%20https://github.com/DARTH-git/dampack}}

\leavevmode\hypertarget{ref-Briggs2012}{}%
\CSLLeftMargin{26. }
\CSLRightInline{Briggs AH, Weinstein MC, Fenwick EAL, Karnon J, Sculpher MJ, Paltiel AD. {Model Parameter Estimation and Uncertainty Analysis: A Report of the ISPOR-SMDM Modeling Good Research Practices Task Force Working Group-6.} Medical Decision Making. 2012 Sep;32(5):722--32. }

\leavevmode\hypertarget{ref-Parmigiani1997}{}%
\CSLLeftMargin{27. }
\CSLRightInline{Parmigiani G, Samsa GP, Ancukiewicz M, Lipscomb J, Hasselblad V, Matchar DB. {Assessing uncertainty in cost-effectiveness analyses: Application to a complex decision model}. Medical Decision Making. 1997;17(4):390--401. }

\leavevmode\hypertarget{ref-Parmigiani2002a}{}%
\CSLLeftMargin{28. }
\CSLRightInline{Parmigiani G. {Measuring uncertainty in complex decision analysis models}. Statistical Methods in Medical Research {[}Internet{]}. 2002;11(6):513--37. Available from: \url{http://smm.sagepub.com/cgi/doi/10.1191/0962280202sm307ra\%5Cnhttp://smm.sagepub.com/content/11/6/513.short}}

\leavevmode\hypertarget{ref-Briggs2002}{}%
\CSLLeftMargin{29. }
\CSLRightInline{Briggs AH, Goeree R, Blackhouse G, O'Brien BJ. {Probabilistic Analysis of Cost-Effectiveness Models: Choosing between Treatment Strategies for Gastroesophageal Reflux Disease}. Medical Decision Making {[}Internet{]}. 2002 Jul;22(4):290--308. Available from: \url{http://mdm.sagepub.com/cgi/doi/10.1177/027298902400448867}}

\leavevmode\hypertarget{ref-Stinnett1998b}{}%
\CSLLeftMargin{30. }
\CSLRightInline{Stinnett AA, Mullahy J. {Net Health Benefits: A New Framework for the Analysis of Uncertainty in Cost-Effectiveness Analysis}. Medical Decision Making {[}Internet{]}. 1998 Apr;18(2):S68--80. Available from: \url{http://mdm.sagepub.com/cgi/doi/10.1177/0272989X9801800209}}

\leavevmode\hypertarget{ref-Alarid-Escudero2019}{}%
\CSLLeftMargin{31. }
\CSLRightInline{Alarid-Escudero F, Enns EA, Kuntz KM, Michaud TL, Jalal H. {"Time Traveling Is Just Too Dangerous" But Some Methods Are Worth Revisiting: The Advantages of Expected Loss Curves Over Cost-Effectiveness Acceptability Curves and Frontier}. Value in Health. 2019;22(5):611--8. }

\leavevmode\hypertarget{ref-Goldhaber-Fiebert2015}{}%
\CSLLeftMargin{32. }
\CSLRightInline{Goldhaber-Fiebert JD, Jalal HJ. {Some Health States Are Better Than Others: Using Health State Rank Order to Improve Probabilistic Analyses}. Medical Decision Making {[}Internet{]}. 2015;36(8):927--40. Available from: \url{http://mdm.sagepub.com/cgi/doi/10.1177/0272989X15605091}}

\leavevmode\hypertarget{ref-Briggs2003}{}%
\CSLLeftMargin{33. }
\CSLRightInline{Briggs AH, Ades AE, Price MJ. {Probabilistic Sensitivity Analysis for Decision Trees with Multiple Branches: Use of the Dirichlet Distribution in a Bayesian Framework}. Medical Decision Making {[}Internet{]}. 2003 Aug;23(4):341--50. Available from: \url{http://mdm.sagepub.com/cgi/doi/10.1177/0272989X03255922}}

\leavevmode\hypertarget{ref-Wickham2021}{}%
\CSLLeftMargin{34. }
\CSLRightInline{Wickham H, Bryan J. {Testing}. In: R packages Organize, test, document and share your code. 2nd ed. Sebastopol, CA: O'Reilly Media; 2021. p. 1--3. }

\leavevmode\hypertarget{ref-Grimmett2014}{}%
\CSLLeftMargin{35. }
\CSLRightInline{Grimmett G, Welsh D. {Markov Chains}. In: Probability: An introduction {[}Internet{]}. 2nd ed. Oxford University Press; 2014. p. 203--3. Available from: \href{https://www.statslab.cam.ac.uk/\%7B~\%7Dgrg/teaching/chapter12.pdf}{www.statslab.cam.ac.uk/\{\textasciitilde\}grg/teaching/chapter12.pdf}}

\leavevmode\hypertarget{ref-Axler2005}{}%
\CSLLeftMargin{36. }
\CSLRightInline{Axler S, Gehring FW, Ribet KA. {Difference Equations}. In New York, NY: Springer; 2005. Available from: \url{http://link.springer.com/10.1007/0-387-27645-9}}

\leavevmode\hypertarget{ref-Cao2016}{}%
\CSLLeftMargin{37. }
\CSLRightInline{Cao Q, Buskens E, Feenstra T, Jaarsma T, Hillege H, Postmus D. {Continuous-Time Semi-Markov Models in Health Economic Decision Making: An Illustrative Example in Heart Failure Disease Management}. Medical Decision Making {[}Internet{]}. 2016;36(1):59--71. Available from: \url{http://mdm.sagepub.com/cgi/doi/10.1177/0272989X15593080}}

\leavevmode\hypertarget{ref-Begun2013}{}%
\CSLLeftMargin{38. }
\CSLRightInline{Begun A, Icks A, Waldeyer R, Landwehr S, Koch M, Giani G. {Identification of a multistate continuous-time nonhomogeneous Markov chain model for patients with decreased renal function.} Medical Decision Making {[}Internet{]}. 2013;33(2):298--306. Available from: \url{http://www.ncbi.nlm.nih.gov/pubmed/23275452}}

\leavevmode\hypertarget{ref-Soares2012}{}%
\CSLLeftMargin{39. }
\CSLRightInline{Soares MO, Canto E Castro L. {Continuous time simulation and discretized models for cost-effectiveness analysis}. PharmacoEconomics {[}Internet{]}. 2012 Dec;30(12):1101--17. Available from: \url{http://www.ncbi.nlm.nih.gov/pubmed/23116289}}

\leavevmode\hypertarget{ref-Frederix2013a}{}%
\CSLLeftMargin{40. }
\CSLRightInline{Frederix GWJ, Hasselt JGC van, Severens JL, Hövels AM, Huitema ADR, Raaijmakers JAM, et al. {Development of a framework for cohort simulation in cost-effectiveness analyses using a multistep ordinary differential equation solver algorithm in R.} Medical Decision Making {[}Internet{]}. 2013;33(6):780--92. Available from: \url{http://www.ncbi.nlm.nih.gov/pubmed/23515213}}

\end{CSLReferences}

\begin{landscape}

\section*{Supplementary Material}

\hypertarget{cohort-tutorial-model-components}{%
\subsection{Cohort tutorial model
components}\label{cohort-tutorial-model-components}}

\hypertarget{table-i}{%
\subsubsection{Table I}\label{table-i}}

This table contains an overview of the key model components used in the
code for the Sick-Sicker example from the
\href{http://darthworkgroup.com/}{DARTH} manuscript:
\href{https://arxiv.org/abs/2001.07824}{``An Introductory Tutorial on Cohort State-Transition Models in R Using a Cost-Effectiveness Analysis Example''}. The first column gives the
mathematical notation for some of the model components that are used in
the equations in the manuscript. The second column gives a description
of the model component with the R name in the third column. The forth
gives the data structure, e.g.~scalar, list, vector, matrix etc, with
the according dimensions of this data structure in the fifth column. The
final column indicated the type of data that is stored in the data
structure, e.g.~numeric (5.2,6.3,7.4), category (A,B,C), integer
(5,6,7), logical (TRUE, FALSE).

\begin{longtable}[]{@{}
  >{\raggedright\arraybackslash}p{(\columnwidth - 10\tabcolsep) * \real{0.10}}
  >{\raggedright\arraybackslash}p{(\columnwidth - 10\tabcolsep) * \real{0.33}}
  >{\raggedright\arraybackslash}p{(\columnwidth - 10\tabcolsep) * \real{0.14}}
  >{\raggedright\arraybackslash}p{(\columnwidth - 10\tabcolsep) * \real{0.14}}
  >{\raggedright\arraybackslash}p{(\columnwidth - 10\tabcolsep) * \real{0.17}}
  >{\raggedright\arraybackslash}p{(\columnwidth - 10\tabcolsep) * \real{0.12}}@{}}
\toprule
Parameter & Description & R name & Data structure & Dimensions & Data
type \\
\midrule
\endhead
\(n_t\) & Time horizon & \texttt{n\_cycles} & scalar & & numeric \\
& Cycle length & \texttt{cycle\_length} & scalar & & numeric \\
\(v_s\) & Names of the health states & \texttt{v\_names\_states} &
vector & \texttt{n\_states} x 1 & character \\
\(n_s\) & Number of health states & \texttt{n\_states} & scalar & &
numeric \\
\(v_{str}\) & Names of the strategies & \texttt{v\_names\_str} & scalar
& & character \\
\(n_{str}\) & Number of strategies & \texttt{n\_str} & scalar & &
character \\
\(d_c\) & Discount rate for costs & \texttt{d\_c} & scalar & &
numeric \\
\(d_e\) & Discount rate for effects & \texttt{d\_e} & scalar & &
numeric \\
\(\mathbf{d_c}\) & Discount weights vector for costs & \texttt{v\_dwc} &
vector & (\texttt{n\_t} x 1 ) + 1 & numeric \\
\(\mathbf{d_e}\) & Discount weights vector for effects & \texttt{v\_dwe}
& vector & (\texttt{n\_t} x 1 ) + 1 & numeric \\
& Sequence of cycle numbers & \texttt{v\_cycles} & vector &
(\texttt{n\_t} x 1 ) + 1 & numeric \\
\(\mathbf{wcc}\) & Within-cycle correction weights & \texttt{v\_wcc} &
vector & (\texttt{n\_t} x 1 ) + 1 & numeric \\
\(age_{_0}\) & Age at baseline & \texttt{n\_age\_init} & scalar & &
numeric \\
\(age\) & Maximum age of follow up & \texttt{n\_age\_max} & scalar & &
numeric \\
\(M\) & Cohort trace & \texttt{m\_M} & matrix & (\texttt{n\_t} + 1) x
\texttt{n\_states} & numeric \\
\(m_0\) & Initial state vector & \texttt{v\_m\_init} & vector & 1 x
\texttt{n\_states} & numeric \\
\(m_t\) & State vector in cycle \(t\) & \texttt{v\_mt} & vector & 1 x
\texttt{n\_states} & numeric \\
& & & & & \\
& \textbf{Transition rates and probabilities} & & & & \\
\(r_{[H,S1]}\) & Constant rate of becoming Sick when Healthy  &
\texttt{r\_HS1} & scalar & & numeric \\
\(r_{[S1,H]}\) & Constant rate of getting Healthy when Sick &
\texttt{r\_S1H} & scalar & & numeric \\
\(r_{[S1,S2]}\) & Constant rate of getting Sicker when Sick &
\texttt{r\_S1S2} & scalar & & numeric \\
\(r_{[S1,S2]_{trtB}}\) & From Sicker to Sick under treatment B
conditional on surviving & \texttt{r\_S1S2\_trtB} & scalar & &
numeric \\
\(r_{[H,D]}\) & Constant rate of dying when Healthy (all-cause mortality
rate) & \texttt{r\_HD} & scalar & & numeric \\
\(r_{[S1,S2]}\) & Constant rate of becoming Sicker when Sick &
\texttt{r\_S1S2} & scalar & & numeric \\
\(r_{[S1,S2]_{trtB}}\) & Constant rate of becoming Sicker when Sick for
treatment B & \texttt{r\_S1S2\_trtB} & scalar & & numeric \\
\(p_{[H,S1]}\) & Probability from Healthy to Sick conditional on surviving &
\texttt{p\_HS1} & scalar & & numeric \\
\(p_{[S1,H]}\) & Probability from Sick to Healthy conditional on surviving &
\texttt{p\_S1H} & scalar & & numeric \\
\(p_{[S1,S2]}\) & Probability from Sick to Sicker conditional on surviving &
\texttt{p\_S1S2} & scalar & & numeric \\
\(p_{[S1,S2]_{trtB}}\) & Probability from Sicker to Sick under treatment B
conditional on surviving & \texttt{p\_S1S2\_trtB} & scalar & &
numeric \\

\(hr_{[S1,H]}\) & Hazard ratio of death in Sick vs Healthy &
\texttt{hr\_S1} & scalar & & numeric \\
\(hr_{[S2,H]}\) & Hazard ratio of death in Sicker vs Healthy &
\texttt{hr\_S2} & scalar & & numeric \\
\(hr_{[S1,S2]_{trtB}}\) & Hazard ratio of becoming Sicker when Sick
under treatment B & \texttt{hr\_S1S2\_trtB} & scalar & & numeric \\
\(P\) & Time-independent transition probability matrix* & \texttt{m\_P}
& matrix & \texttt{n\_states} x \texttt{n\_states} & numeric \\
& * \texttt{\_trtX} is used to specify for which strategy the transition
probability matrix is & & & & \\
& & & & & \\
& \textbf{Annual costs} & & & & \\
& Healthy individuals & \texttt{c\_H} & scalar & & numeric \\
& Sick individuals in Sick & \texttt{c\_S1} & scalar & & numeric \\
& Sick individuals in Sicker & \texttt{c\_S2} & scalar & & numeric \\
& Dead individuals & \texttt{c\_D} & scalar & & numeric \\
& Additional costs treatment A & \texttt{c\_trtA} & scalar & &
numeric \\
& Additional costs treatment B & \texttt{c\_trtB} & scalar & &
numeric \\
& Vector of state costs for a strategy & \texttt{v\_c\_str} & vector & 1
x \texttt{n\_states} & numeric \\
& list that stores the vectors of state costs for each strategy &
\texttt{l\_c} & list & & numeric \\
& & & & & \\
& \textbf{Utility weights} & & & & \\
& Healthy individuals & \texttt{u\_H} & scalar & & numeric \\
& Sick individuals in Sick & \texttt{u\_S1} & scalar & & numeric \\
& Sick individuals in Sicker & \texttt{u\_S2} & scalar & & numeric \\
& Dead individuals & \texttt{u\_D} & scalar & & numeric \\
& Treated with treatment A & \texttt{u\_trtA} & scalar & & numeric \\
& Vector of state utilities for a strategy & \texttt{v\_u\_str} & vector
& 1 x \texttt{n\_states} & numeric \\
& List that stores the vectors of state utilities for each strategy &
\texttt{l\_u} & list & & numeric \\
& & & & & \\
& \textbf{Outcome structures} & & & & \\
& Expected QALYs per cycle under a strategy & \texttt{v\_qaly\_str} &
vector & 1 x (\texttt{n\_t} + 1) & numeric \\
& Expected costs per cycle under a strategy & \texttt{v\_cost\_str} &
vector & 1 x (\texttt{n\_t} + 1) & numeric \\
& Vector of expected discounted QALYs for each strategy &
\texttt{v\_tot\_qaly} & vector & 1 x \texttt{n\_states} & numeric \\
& Vector of expected discounted costs for each strategy &
\texttt{v\_tot\_cost} & vector & 1 x \texttt{n\_states} & numeric \\
& Summary matrix with costs and QALYS per strategy &
\texttt{m\_outcomes} & table & \texttt{n\_states} x 2 & \\
& Summary of the model outcomes & \texttt{df\_cea} & data frame & & \\
& Summary of the model outcomes & \texttt{table\_cea} & table & & \\
& & & & & \\
& \textbf{Probabilistic analysis structures} & & & & \\
& Number of PSA iterations & \texttt{n\_sim} & scalar & & numeric \\
& List that stores all the values of the input parameters &
\texttt{l\_params\_all} & list & & numeric \\
& Data frame with the parameter values for each PSA iteration &
\texttt{df\_psa\_input} & data frame & & numeric \\
& Vector with the names of all the input parameters &
\texttt{v\_names\_params} & vector & & character \\
& List with the model outcomes of the PSA for all strategies &
\texttt{l\_psa} & list & & numeric \\
& Vector with a sequence of relevant willingness-to-pay values &
\texttt{v\_wtp} & vector & & numeric \\
& Data frame to store expected costs and effects for each strategy from
the PSA & \texttt{df\_out\_ce\_psa} & data frame & & numeric \\
& Data frame to store incremental cost-effectiveness ratios (ICERs) from
the PSA & \texttt{df\_cea\_psa} & data frame & & numeric \\
& For more details about the PSA structures read \texttt{dampack}'s
vignettes & & & & \\
\bottomrule
\end{longtable}

\end{landscape}
\end{document}